\documentclass[nofootinbib]{revtex4}
\usepackage{amsfonts,amssymb,amsmath,bbm}
\usepackage{mathrsfs}
\usepackage[all]{xy}  % xy diagrams
\usepackage{tikz}  

\newcommand{\threej}[6]{ \left( \begin{array}{ccc} 
 #1 & #2 & #3\\
#4 & #5 & #6 
\end{array}
\right)}
\newcommand{\sixj}[6]{ \left\{ \begin{array}{ccc} 
 #1 & #2 & #3\\
#4 & #5 & #6 
\end{array}
\right\}}
\newcommand{\lj}[3]{\left( \begin{array}{c}  #1  \\#2 \, \,\, #3 \end{array} \right)}
\newcommand{\eff}{\mathrm{eff}}
\newcommand{\Pino}{ F. V.}
\newcommand{\Ignazio}{ G. S. }
\newcommand{\hpsi}{\hat{\Psi}}
\newcommand{\xxx}{\mathrm{x}}

\newcommand{\SLC}{\mathrm{SL}(2,\mathbb{C})}
\newcommand{\SU}{\mathrm{SU}(2)}
\newcommand{\ie}{\textit{i.e.}~}
\newcommand{\eg}{\textit{e.g.}~}
\newcommand{\Tr}{\mathrm{Tr}}
\begin{document}
\begin{flushright}
Preprint AEI-2010-157
\end{flushright}

\title{Towards classical geometrodynamics from Group Field Theory hydrodynamics}
\author{\bf Daniele Oriti\footnote{email: doriti@aei.mpg.de}} 
\author{\bf Lorenzo Sindoni\footnote{email: sindoni@aei.mpg.de}}
\affiliation{Max-Planck-Institut f\"ur Gravitationsphysik\\
Albert Einstein Institute\\
Am M\"uhlenberg 2, 14476, Golm, Germany, EU}

\begin{abstract}
We take the first steps towards identifying the hydrodynamics of group field theories (GFTs) and relating this hydrodynamic regime to classical geometrodynamics of continuum space. We apply to GFT mean field theory techniques borrowed from the theory of Bose condensates, alongside standard GFT and spin foam techniques. The mean field configuration we study is, in turn, obtained from loop quantum gravity coherent states. We work in the context of 2d and 3d GFT models, in euclidean signature, both ordinary and colored, as examples of a procedure that has a more general validity. We also extract the effective dynamics of the system around the mean field configurations, and discuss the role of GFT symmetries in going from microscopic to effective dynamics. In the process, we obtain additional insights on the GFT formalism itself.

\end{abstract}
\maketitle

\section*{Introduction}
Many approaches to Quantum Gravity have been developed and flourished in recent years \cite{libro}. Some of them build directly upon traditional quantization strategies (Dirac canonical programme, covariant perturbative quantization, sum-over histories or path integral quantization) applied to the gravitational field; examples of this are canonical loop quantum gravity and spin foam models in the continuum formulation \cite{thomas, carlo} or the Asymptotic safety programme. Others rests on traditional quantization methods, usually path integrals, but introduce new ingredients or starting assumptions, for example discrete space(time) structures interpreted either as mere regularization tools or as somehow \lq\lq physical\rq\rq; examples are simplicial approaches like quantum Regge calculus \cite{RC} and (causal) dynamical triangulations \cite{DT}, or spin foam models based on simplicial lattices \cite{alex,review}. String theory \cite{ST} is another example, as it starts with a conventional covariant perturbative quantization strategy as applied to strings, and thus to their graviton excitations, but introduces a pletora of additional structures along the way, including extra dimensions, supersymmetry, and of course the abandonment of the point-like and thus strictly local nature of physical systems and their interactions. Other approaches still take their very start from some new radical hypothesis about the microscopic structure of quantum space and its constituents, variously inspired or motivated from General Relativity and geometrodynamics \cite{CS}, or from more philosophical or mathematical considerations \cite{topos, NCG}. 

This paper deals with  the approach to Quantum Gravity usually called Group Field Theory \cite{iogft, laurentgft, iogft2, gftreview}. It is best understood as a generalization of matrix models for 2d quantum gravity \cite{mm}. Matrix models, defined in terms of pre-geometric, \lq\lq space-free\rq\rq constituents (matrices) with a dynamics dictated by purely combinatorial and discrete (pre-)geometric considerations, generate a sum over 2-dimensional simplicial complexes in their perturbative Feynman expansion, later to become a sum over smooth random surfaces in an appropriate continuum limit (understood in terms of a phase transition of the pre-geometric system), defining a path integral for 2d quantum gravity. In a similar way group field theory models are defined in terms of \lq\lq pre-geometric\rq\rq, \lq\lq space-free\rq\rq fundamental entities, fields over several copies of group manifolds or their Lie algebras, with a dynamics again dictated by purely combinatorial, discrete (pre-)geometric inputs, and generate d-dimensonal simplicial complexes in their Feynman expansion. Their historic origin from work in simplicial quantum gravity, state sum models and, more recently, in loop quantum gravity and spin foam models is manifest in the very mathematical structures they are based on. More precisely, while the combinatorics of field arguments in the GFT action is dictated, as in matrix models, by the requirement that the corresponding Feynman  diagrams have the combinatorial structure of simplicial complexes of the appropriate dimension, the arguments of the field themselves and the specific form of the GFT action are motivated by results in loop quantum gravity and discrete gravity (and non-commutative geometry). As a result the GFT Feynman amplitudes take generically the form of simplicial gravity path integrals \cite{ioaristide} or, equivalently, spin foam models \cite{carlomike}, i.e. sum-over-histories of spin networks. The GFT field itself can be interpreted as a second quantized simplex or spin network vertex wavefunction. We refer to the reviews \cite{iogft,iogft2,laurentgft,gftreview} for more details. 

Most of the above approaches, including some of those starting as straightforward quantizations of continuum General Relativity, end up identifying purely algebraic and discrete structures as fundamental building blocks (states) of quantum space, each characterized by a finite number of quantum degrees of freedom. However convincing the proposed microscopic dynamics for such fundamental building blocks is, the main task that all such approaches have to fulfill becomes that of showing how the algebraic, discrete quantum degrees of freedom give rise to a continuum geometric description of spacetime with its dynamics governed by General Relativity, in some approximation. 

Various solution to this problem have been proposed. These range from coherent states in canonical loop gravity \cite{coherentLQG} to statistical field theory methods in simplicial quantum gravity \cite{RC,DT} and in other
discrete pre-geometric models of space \cite{fotini}, to ideas coming from quantum information theory \cite{fotini2}. Unless a suitable interpretation in terms of continuum geometries is found for both the discrete quantum structures  and their microscopic dynamics as such (for attempts in this direction see \cite{carlograviton, graviton2,SFcosmology}), a continuum space is likely to be understood as built out of a very large number of microscopic quantum building blocks, and the continuum limit of the microscopic dynamics  arises in a thermodynamic limit, and is thus possibly subject to critical phenomena. This is the idea at the root of simplicial gravity approaches, hence the use of statistical field theory methods, but also the origin of the very difficulties involved in tackling the problem of the continuum, since one has necessarily to deal with very large numbers of quantum degrees of freedom and their collective behaviour. The same difficulty is encountered in the context of loop quantum gravity coherent states and spin foam models, even more so if dynamical spatial graphs or spacetime lattices are involved (as the graph-changing nature of the Hamiltonian constraint operator in loop quantum gravity would seem to require). 

At this point, group field theories enter the stage as the framework in which, on the one hand, the very same fundamental structures identified by other approaches as building blocks of space (simplices or spin networks) can be treated in a unified way, with all the techniques already developed in those approaches. On the other hand, new techniques become available, thanks to the GFT embedding. In particular, GFT become the natural formalism for studying the physics of a large number of the same quanta of space identified in other approaches, because it is a second quantized formalism of the same, well-adopted to the study of many-particle systems. Let us clarify this point further, as it is central to the work presented in this paper. 

Once understood that the same building blocks of space used in other approaches arise as quanta of the GFT field, and that the GFT action defines a microscopic dynamics for them, then the study of their effective dynamics (classical or quantum) in a thermodynamic limit becomes a problem in statistical GFT. Quantum space is then interpreted, at least at the level of analogy, as a sort of exotic condensed matter system, a condensate/fluid made out of large numbers of GFT quanta, to be studied with tools and ideas from condensed matter theory. 
The continuum and semi-classical approximation that leads from microscopic quantum structures to smooth continuum space(time) can then be seen as analogous to the  hydrodynamic approximation that leads from the microscopic quantum description of a few atoms system to the hydrodynamic description of a classical fluid made out of large numbers of the same atoms. The GFT itself becomes then the quantum gravity analog of the microscopic quantum field theory of non-relativistic atoms that underlies any condensed matter system, and any (quantum or classical) fluid. Classical geometrodynamics (including General Relativity) should then arise, in a way to be understood, from GFT hydrodynamics, in some appropriate phase of the theory, also to be identified. It is unlikely, in fact, that the infinite number of GFT degrees of freedom that are necessary to obtain a smooth spacetime and its relativistic dynamics will organize themselves in a single macroscopic phase, and thus that no understanding of GFT phase transitions will be needed to understand the emergence of continuum classical spacetime from GFT. This perspective has been argued for in \cite{gftfluid}. 

The idea of spacetime as a fluid/condensate has of course been put forward repeatedly \cite{hu}, and is somehow the conceptual underpinning of condensed matter analog gravity models \cite{analogreview, volovik}, together with the idea of General Relativity as the emergent hydrodynamics or thermodynamics of microscopic pre-geometric building blocks \cite{GRhydro, jacobson,eling,Eling2,GRthermo,goffredo}. The body of this work (in particular the one on quantum fluids\footnote{For concrete implementations of these ideas within idealized BEC models, and extended discussions, see \cite{BECdyn,lor}.}), we believe, represents an important guide for the extraction of geometrodynamics from GFT. Conversely, we also believe that GFTs can represent a candidate for the microscopic description of the \lq\lq atoms of quantum space\rq\rq that all this work somehow hints at.   

Once accepted, at least provisionally, the above perspective, the task is to identify the best strategy and mathematical tools to tackle the problem of the continuum in a GFT context. Being a problem in statistical GFT, obviously the renormalization group is a key asset. Indeed a programme of GFT renormalization has recently started \cite{GFTrenorm}, which has extracting the continuum limit of GFT as its main physical goal, beyond the many mathematical insights it is providing \cite{Razvan2}. Another key method for extracting information about the phase structure of a condensed matter system, and for the extraction of the corresponding effective dynamics, is mean field theory.  In particular, mean field theory is the most direct route from the microscopic atomic dynamics to the effective hydrodynamics in the case of quantum fluids and Bose condensates, which are the systems we tentatively use as paradigms of what  the emergence of spacetime from quantum gravity may entail. 

The application of mean field theory requires first of all the identification of a candidate macroscopic configuration of the system (and associated collective variables), whose dynamics encodes the collective dynamics of the many-particles constituting it. Also, the fluctuations around such new vacuum become the relevant  (once more, collective) degrees of freedom at the macroscopic level, with an associated dynamics in general very different from the microscopic one governing the underlying atoms. All this works under the assumption that this new vacuum will be very different from the microscopic quantum vacuum of the system (the Fock vacuum), and that the physics one is looking for at the macroscopic level is best understood close to the assumed macroscopic configuration, rather than the microscopic Fock vacuum. This a particularly reasonable assumption in the GFT context,  where the vacuum  around which the  usual perturbative expansion takes place is interpreted physically as a \lq\lq no space\rq\rq state \cite{iogft,iogft2,laurentgft,gftreview}, in  which no space structure at all exists, neither topological nor geometrical. The same reasoning suggests that the usual spin foam models or the equivalent simplicial path integrals are not the most convenient definitions of the dynamics of the degrees of freedom that correspond to a smooth spacetime and its geometry, because they arise as Feynman amplitudes of GFT models {\it around the no-space vacuum} \cite{gftfluid}. It suggests to look instead for an effective dynamics of perturbations around a different vacuum. The identification of the relevant macroscopic vacuum and of the corresponding dynamics is however a highly non-trivial task, in condensed matter and, even more, in our quantum gravity context. {The best one can do, usually, is to proceed by educated guesses, and then test the resulting hydrodynamic theory against observation. In quantum gravity we cannot (yet) make use of experimental inputs, but we can still proceed using intuition and the large amount of theoretical ideas accumulated so far, and study the formal aspects of the effective theory obtained as result of our hypothesis.}

This is what we do in this, largely exploratory, paper. We move the first steps towards establishing a hydrodynamic limit of GFTs and in relating this hydrodynamic regime to classical geometrodynamics of continuum space. First of all, this requires the identification of a candidate (non-perturbative)  macroscopic vacuum. Here we start from the results obtained in the context of semiclassical loop quantum gravity, in particular on LQG coherent states. We identify, from a simple analysis of LQG coherent states, a candidate coherent state wave function associated to vertices of spin networks. This coherent state wave function will depend on parameters that can be interpreted in geometric terms, following again the LQG results, and that play the role of {\it order parameters} in the GFT context. We identify this candidate vacuum state in section III, together with its geometric interpretation. Then, we re-interpret the same wave function as a classical GFT field, and use it as candidate mean field configuration for the GFT dynamics. We do so in section IV and V, considering in detail the simpler cases of GFT models for quantum BF theory in 2d and for 3d quantum gravity, in euclidean signature, and for both ordinary and colored GFTs. We extract the mean field theory equations for the order parameters, as resulting from the GFT dynamics. These hydrodynamic equations have a geometrodynamic interpretation, that we elucidate to some extent. We are not able, at this stage, to give a complete and transparent geometric re-writing of these equations, nor to obtain a precise mapping of these equations with those of General Relativity. This is perhaps the main limitation of our results. Still, we believe we open up a new interesting direction for further investigations. Having obtained the mean field or hydrodynamic equations for the (geometric) order parameters characterizing this GFT vacuum, we move on to extract the effective dynamics for perturbations around it, in section VI. The main purpose, here, is first of all to illustrate the general procedure and its main features; second, it is meant to show how, concretely, the effective dynamics could, on the one hand, differ from the microscopic one and, on the other hand, depend on it, in the spin foam formulation. A more detailed analysis that the one we perform, once more, is certainly needed, but most important, what is needed is a careful study of the physical interpretation of the perturbation field itself, and of the degrees of freedom it carries. At the present stage, it is unclear if these should be interpreted as matter fields\footnote{Recent work in the GFT context \cite{eterawinston, ioeteraflorian} has applied a similar strategy to the one we follow in this paper, and can also be understood from a condensed matter analogy \cite{emergentmatterGFT}. The purpose of such work was, in fact, the extraction of effective (non-commutative) matter field theories as perturbations around exact solutions of the GFT equations of motion. No effective dynamics for the background GFT configurations chosen was investigated.} living on the geometric background defined by the mean GFT field, or if they have a geometric interpretation as well, so that, for example, the full geometric character of the effective GFT hydrodynamics  is to be looked for in the coupled equations for order parameters {\it and} quasi-particle fluctuations around the mean field. We leave this issue for future work. Last, in section VII, we discuss the issue of GFT symmetries, in particular diffeomorphisms, from the point of view of the application of mean field theory methods. A summary of our results, and some outlook, are provided in the last section.

To clarify the path we follow and the conceptual framework for the results we present from section III onwards, it will be useful to briefly review how the mean field theory method is applied in a very well understood physical system, namely Bose--Einstein condensates, to extract the effective hydrodynamics of the condensate. We do so in section II, where we also point out the main differences (as well as the similarities) between our approach in the GFT context and the case of Bose condensates, together with its limitations.  

\section{Mean field theory and effective hydrodynamics in Bose condensates: analogies and differences with the GFT case}

The paradigm we will follow is mean field theory applied to the study of Bose--Einstein condensates. For an extensive discussion on the subject, we refer to \cite{BEC}. Here, we recall only the main features relevant for our purposes. 

We will focus on the particular case of Bose--Einstein condensation in dilute gases. For these systems, the most convenient treatment is provided by the second quantized formalism, based on quantum field operators
\begin{equation}
\hpsi(\xxx) = \sum_{i} \hat{a}_{i} u_{i}(\xxx),
\end{equation}
where $i$ is an index labelling an orthonormal basis of (single particle) wavefunctions $u_{i}$, with
\begin{equation}
\int_{\mathcal{V}} u_{i}(\xxx)u_{j}^{*}(\xxx) d^{3}x = \delta_{ij}, 
\end{equation}
and $\hat{a}_{i}$ are (bosonic) annihilation operators, obeying
\begin{equation}
[\hat{a}_{i},\hat{a}_{j}] = 0; \qquad [\hat{a}_{i},\hat{a}^{\dagger}_{j}] = \delta_{ij}. 
\end{equation}
These operators are associated to the creation/annihilation of fundamental particles (atoms) of the system. 
The typical Hamiltonian describing the dynamics of a dilute gas of bosons (in the local approximation) has the form:
\begin{equation} \label{Ham}
\hat{H} = \int 
\hpsi^{\dagger}(\xxx) \left( -\frac{\hbar^{2}\nabla^{2}}{2m} - \mu + V(\xxx) + \frac{\kappa}{2} \hpsi^{\dagger}(\xxx) \hpsi(\xxx) \right) \hpsi(\xxx)
\,d^{3}x,
\end{equation}
where $m$ is the mass of the bosons, $V(\xxx)$ is an external potential used to confine the system in a given region of space ($\mathcal{V}$), $\kappa$ is a (positive) constant encoding the strength of two-particle interactions and $\mu$ is a chemical potential, playing the role of a Lagrange multiplier used to fix the total number of particles. The above Hamiltonian corresponds to the case of a single atomic species, but the formalism can be immediately extended (2BEC, spinor BEC, etc).

The main problem is then the correct identification of the ground state. If the system were non-interacting, the ground state, below the critical temperature, would be the state in which all bosons occupy the same single particle ground (lowest energy) state (Bode condensate). The case of interacting particles is much more involved. However, one can make a guess at the macroscopic properties of the system by assuming that, below a certain temperature (to be computed), the condensation does take place even in the interacting case, and even for repulsive atomic interactions (positive $\kappa$). 
The simplest way to implement this idea is to assume that the (interacting) ground state (G.S.) is a state in the Fock space, $|\Ignazio \rangle $,
such that the operator $\hpsi(\xxx)$ acts on it as multiplication by a function,
\begin{equation}
\hpsi(\xxx) | \Ignazio \rangle \approx \psi(\xxx) | \Ignazio \rangle 
\end{equation}
Clearly, this kind of state is crucially different from the Fock vacuum (F.V.) $|\Pino\rangle$, for which $\hpsi(\xxx) | \Pino \rangle = 0$. 

A simple example of state possessing this property is a (second quantized) coherent state. Consider the state:
\begin{equation}
|z_{i}\rangle = e^{-|z|^{2}/2} \exp(z_{i} \hat{a}_{i}^{\dagger}) |\Pino\rangle
\end{equation}
It is immediate to see that this state is an eigenstate of the field operator $\hpsi(\xxx)$:
\begin{equation}
\hpsi(\xxx)|z_{i}\rangle = z u_{i}(\xxx) |z_{i}\rangle = \hpsi(x) | \Ignazio \rangle.
\end{equation}
The content of the state is characterized by the expectation value of the number operator of each mode:
\begin{equation}
n_{j}(i) = \langle z_{i} | \hat{a}_{j}^{\dagger} \hat{a}_{j}| z_{i} \rangle = \delta_{ij} |z_{i}|^{2}
\end{equation}

Therefore, as a working hypothesis, one {\it assumes} that the ground state of the many body system is 
a coherent state, with a macroscopic occupation number of a suitably chosen single particle state: $| \Ignazio \rangle = | z_i \rangle$. Concretely, this means that
one is working in a regime in which the field operator $\hpsi(\xxx)$ can be effectively separated as
\begin{equation}
\hpsi(\xxx) \approx \psi(\xxx) \mathbb{I} + \hat{\chi}(\xxx),
\end{equation}
where $\psi(\xxx)$, often called the condensate wavefunction, encodes the information about the ground state, while $\hat{\chi}$ encodes the effect
of deviations from the mean field $\psi$. This splitting must then be introduced into the equations of motion for the field operator $\hpsi$. In turn,
this equation of motion will become an equation for the classical field $\psi$, including the effects of perturbations $\hat{\chi}$. The logic is clear: the mean
field $\psi$ must be determined self-consistently from the mean field equations. 
At the lowest order, neglecting the fluctuations $\hat{\chi}$, this equation assumes the form
\begin{equation}
i\hbar\frac{\partial}{\partial t} \psi = -\frac{\hbar^{2} \nabla^{2}}{2m} \psi - \mu \psi + V(\xxx) \psi + \kappa |\psi|^{2}\psi , 
\end{equation}
known as the Gross--Pitaevski hydrodynamic equation. Its solution provides the description of the macroscopic properties of the condensate described by the wavefunction $\psi$. It can be re-written in terms of two real fluid equations (Euler and continuity) for a perfect fluid of density $\rho$ and velocity $\vec{v}$ such that
\begin{equation}
\psi = \sqrt{\rho} e^{-i \theta},\qquad \vec{v} \propto \nabla\theta.
\end{equation}
It is in this sense that the mean field approximation gives immediately the hydrodynamic description of the system.

It is important to keep in mind that this method must be self-consistent: we have started assuming that the ground state is a condensed state, in which many particles condense in the same single-particle state. However, it is only after that we have found the solution of the GP equation, with the appropriate boundary conditions, and we have established that this solution corresponds to a configuration in which a large fraction of the $N$ bosons originally present in the system is occupying the condensed state\footnote{Explicitly we must check that $\int d^{3}x |\psi(\xxx)|^{2} \approx N$.} that we can say that our system condenses. Also, even if the method proves to be consistent, this mean field treatment is just an approximation that takes into account only the field configuration on which the coherent state is peaked. The spread of the coherent state itself is neglected (this corresponds to neglecting terms containing, e.g. $\langle \hat{\chi}^{2}\rangle$). In general, more refined tools are needed \cite{castindum}.
Nonetheless, the rough description in terms of the condensate wavefunction allows to understand a number of interesting features of the physics of the system. From a perspective {\it \'a la} Landau, the condensate wave function is an order parameter describing the phase transition. It plays two roles. First of all, it contains the basic information about the hydrodynamical properties of the condensed fraction (density, velocity profile etc.).
Second, it determines the symmetries and the general properties of the effective dynamics of perturbations around the condensed state itself. In fact, if we consider the dynamics of phonons, the elementary excitations above the condensate, their various properties (internal symmetries, dispersion relations and spacetime symmetries, etc.)
are essentially controlled by the wavefunction $\psi$.

\

It is useful to make a comparison with the GFT case. In the BEC phenomenon the ground state being selected has the peculiarity of peaking the second quantized field operator on a given classical field configuration. From the point of view of the particle content, it is a superposition of many particle states with different occupation numbers. Finally, the mean field represents the order parameter that describes the condensation, and encodes the macroscopic hydrodynamic variables that effectively describe the semiclassical state.

In the case of GFTs, the second quantization formalism, with a definition of a Fock space and a corresponding many particle interpretation is still not understood. Therefore, a proper GFT analogue of the BEC mechanism is not available either. Consequently, in what follows, we will be still working in a first quantized scenario, corresponding to the kinematical Hilbert space of Loop Quantum Gravity, whose states appear as boundary data in GFT transition amplitudes, and therefore the analogy with the BEC case must be taken with care. The candidate ground state will be obtained from LQG coherent states, and correspond to wavefunctions for spin network vertices (the GFT quanta, i.e. their candidate \lq atoms of quantum space\rq) characterized by (peaked on) $\SLC$ group elements, defining points in the classical phase space appropriate for the gauge theories considered. The situation is similar to a BEC mean field approximation in which one assumes that the configuration $u_{0}(\xxx)$ is a wavefunction peaked on a certain point of phase space for a single particle. This is clearly a much stronger semi-classicality condition than only requiring the quantum field operator acquires a nonzero vacuum expectation value, as in the true BEC case. 

Still, within the limits of our analysis, the two mean field approximations in BEC and GFT (as performed in this paper) share the same philosophy. The group elements where the peaks are situated play the role of order parameters, as the condensate wavefunction does. Also, they will be dynamically determined by equations that will be derived from the GFT equations of motion in the same way in which the Gross--Pitaevski equation is derived from the non-linear Schr\"odinger equation for second quantized field operator. As it has been discussed in details in the literature, the $\SLC$ group elements determine the mean value of certain geometric quantum operators, and correspond to classical geometric fields, while the GFT dynamics is expected to encode the full quantum dynamics of the microscopic degrees of freedom of quantum space, and the equations that we derive for them represent a form of geometrodynamics, here derived from the microscopic dynamics of quantum space as encoded in a GFT, in a mean field (hydrodynamic) approximation.

\section{From LQG coherent states to an ansatz for for a GFT background configuration}

We now introduce briefly kinematical loop quantum gravity (LQG) coherent states, and then show how to extract from them a candidate background configuration to be used in the mean field approximation of GFT models. More precisely, we show how LQG coherent states can be obtained by convolution of vertex wave functions, also characterized, as the LQG coherent states, by $\SLC$ group elements, and ready to be used as mean GFT fields, as we do in the following.

\subsection{Coherent states on $\SU$ and vertex wavefunctions}
(Complexifier) coherent states were introduced within Loop Quantum Gravity in order to investigate the semiclassical limit of the theory. They are wavefunctions associated to graphs, which have the properties to be peaked around certain values of the classical phase space of the theory, parametrized by holonomies (group elements of $\SU$) and fluxes (elements of the Lie algebra $\mathfrak{su}(2)$). 
The expectation values of (polynomial functions of) these operators coincides with the classical values at the corresponding phase space point. Also, they minimize uncertainties for a specific set of kinematical observables and within a given class of quantum states. Given this nice kinematical properties, a key question becomes then to what extent this class of states satisfies the quantum dynamics of the theory, and whether the same dynamics preserve their semi-classical properties. While we will not address this question within the LQG canonical framework, we will in fact study the issue of whether the associated vertex wave functions, our candidate GFT mean field, solves the GFT equations of motions.

For a complete discussion of their properties and their relevance for the semiclassical approximation within LQG, we refer to \cite{coherentLQG}. Here we summarize only their definition.

Let $\gamma$ be a graph, and let $\vec{g}$ denote a set of group elements ($\SU$) associated to its links (one per link); these group elements should be considered as parameters defining the state. The coherent states on $\gamma$, in the connection representation, are functions from $\SU^{E(\gamma)}$ to the complex numbers, given by:
\begin{equation}
\Psi[\gamma,\vec{g};t](\vec{h}) := \int (d\tilde{g})^{|V(\gamma)|} \psi[\gamma,\vec{g};t](\{ \tilde{g}(e(1))h_{e} \tilde{g}^{-1}(e(0))\}), 
\end{equation}
where $\vec{h}$ denotes the set of the group variables (one per link), the integrations over the group associated to the vertices $V(\gamma)$ of the graph $\gamma$ enforce gauge invariance (Gauss constraint), and where the gauge-variant wavefunction,
\begin{equation}
\psi[\gamma,\vec{g};t](\vec{h}) = \prod_{e\in E(\gamma)} \rho_{t}(h_{e}g_{e}^{-1}),
\end{equation}
is a product of heat kernels ($\rho_{t}$) on the group manifold $\SU$ \cite{camporesi} with spread (diffusion time) $t$. 
Each heat kernel is a function of the group element $h$ peaked on the element $g$ with spread $t$.

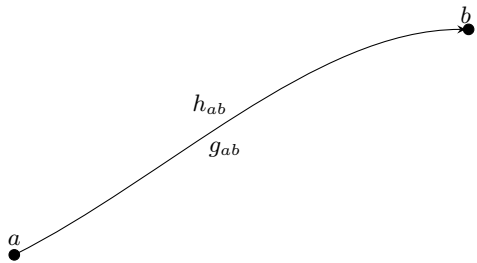
\begin{figure}
\begin{tikzpicture}[scale=2, >=stealth]
\draw[->] (0,0) .. controls (1,0.5) and (2,1.5) .. (3,1.5);
\filldraw (0,0) [black] circle (1pt);
\filldraw (3.02,1.5) [black] circle (1pt);
\path (0,0.1) node {$a$};
\path (3,1.6) node {$b$};
\path (1.3,1) node {$h_{ab}$};
\path (1.4,0.7) node {$g_{ab}$};
\end{tikzpicture}
\caption{Entire oriented link from $a$ to $b$, with associated group variable $h_{ab}$ and peak group element $g_{ab}$ of the heat kernel.}
\label{entirelink}
\end{figure}

\begin{figure}
\begin{tikzpicture}[scale=2, >=stealth]
\draw[->, dashed] (0,0) .. controls (1,0.5) and (2,1.5) .. (3,1.5);
\filldraw (0,0) [black] circle (1pt);
\filldraw (3.02,1.5) [black] circle (1pt);
\path (0,0.1) node {$a$};
\path (3,1.6) node {$b$};
\path (1.15,1.05) node {$h_{ab}$};
\path (1.4,0.5) node {$g_{ab}$};
\draw [thick] (1.3,1) -- (1.6,0.8);
\draw [thick] (1.25,0.95) -- (1.55,0.75);
\filldraw (1.412,0.88) [gray] circle (1pt);
\draw[->] (0,0) .. controls (0.7,0.45) .. (1.35, 0.87);
\draw[->] (3,1.5) .. controls (2.7,1.45) and (2,1.25) .. (1.47, 0.89);
\path (0.5,0.5) node {$h_{a}$};
\path (0.7,0.2) node {$g_{a}$};
\path (2.3,1.5) node {$h_{b}$};
\path (2.5,1.2) node {$g_{b}$};
\path (1.5,1) node {$P$};
\end{tikzpicture}
\caption{Splitting the link $a$ to $b$. The group elements $h_{a},h_{b},g_{a},g_{b}$ refer to the semi-link from the vertices $a,b$ to $P$, with the appropriate orientation.}
\label{splittinglink}
\end{figure}
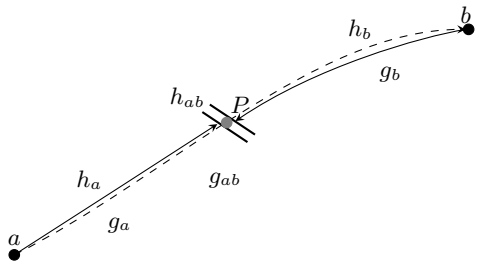

The same states can be rewritten in terms of functions associated to vertices, as we now show.
For each link $(ab)$ as in figure \ref{entirelink}, introduce the intermediate point $P$ and double the associated group elements (see figure \ref{splittinglink}) as:
\begin{equation}
h_{ab}= h_{b}^{-1} h_{a}; \qquad g_{ab} = g_{b}^{-1} g_{a},
\end{equation}
thus artificially doubling the number of group elements associated to the same link (this is not a physical doubling of variables, since the two group elements only enter the expression as the products $h_{ab},g_{ab}$).
We then use the convolution property:
\begin{equation}
\rho_{2t} (g) = \int dh \, \rho_{t_{1}}(gh^{-1})\rho_{t_{2}}(h); \qquad t_{1}+t_{2}=2t.
\end{equation}
For simplicity (but without loss of generality\footnote{Furthermore,
the parameter $t$ should be determined by the quantum dynamics (\eg the Hamiltonian constraint), and a natural assumption is that the
dynamics will select a value of $t$ which will not depend on the portion of the particular graph we are considering, at least in the case in which we will not deal with too inhomogeneous states.}) we will consider $t_{1}=t_{2}=t$.

Neglecting for the moment the group averaging procedure enforcing gauge invariance at each vertex, we obtain:

\begin{equation}
\star = \rho_{2t} (h_{ab}g^{-1}_{ab}) = \rho_{2t} (h_{b}^{-1}h_{a}g_{a}^{-1}g_{b}) = \rho_{2t}(h_{a}g_{a}^{-1}g_{b}h_{b}^{-1}),
\end{equation}
where we have used the property $\rho_{t}(hg)=\rho_{t}(gh)$. Then, we can split the single link by using the convolution property of the heat kernel:
\begin{equation}
\star = \int dq_{ab} \, \rho_{t}(h_{a}g_{a}^{-1} q_{ab}^{-1}) \rho_{t} (q_{ab} g_{b} h_{b}^{-1}) = \int dq_{ab} \, \rho_{t}(h_{a}g_{a}^{-1} q_{ab}^{-1}) \rho_{t} ( h_{b} g_{b}^{-1} q_{ab}^{-1}),
\end{equation}
where we have used the property $\rho_{t}(g^{-1}) = \rho_{t}(g)$. Notice the identical functional dependence on the various group elements in each of the two functions associated to each (oriented) semi-link.

With an obvious change of notation ($ab \rightarrow e$, $a \rightarrow e(0)$ and $b\rightarrow e(1)$), the complete function associated to the graph reads:
\begin{equation}
\psi[\gamma,\vec{g};2t](\vec{h}) = \int (dq)^{|E(\gamma)|} \prod_{e \in E(\gamma)} \rho_{t}(h_{e(0)}g_{e(0)}^{-1}q_{e}^{-1})
\rho_{t}(h_{e(1)}g_{e(1)}^{-1}q_{e}^{-1}).
\end{equation}
This expression can be reorganized as
\begin{equation}
\psi[\gamma,\vec{g};2t](\vec{h}) = \int (dq)^{|E(\gamma)|} \prod_{v \in V(\gamma)} \Phi_{v} (h_{v,{1}}g_{v,{1}}^{-1} q_{e_{v,{1}}}^{-1}; h_{v,{2}}g_{v,{2}}^{-1} q_{e_{v,{2}}}^{-1};...;h_{v,{m_{v}}}g_{v,m_{v}}^{-1} q_{e_{v,m_{v}}}^{-1})
\end{equation}
where $h_{v,i}$ is the $i-$th element associated to the $i-$th link ending in the vertex $v$, $m_{v}$ is the valence of the vertex $v$, and:
\begin{equation}
\Phi_{v} (g_{1}; ...;g_{m_{v}}) = \prod_{i=1}^{m_{v}} \rho_{t} (g_{i})
\end{equation}
This vertex function is a function going from $G^{m_{v}} \rightarrow \mathbb{R} $, thus can be interpreted as a GFT field. Correspondingly, the whole wavefunction can be rewritten in terms of a suitable operator defined in a group field theory\footnote{Also the case in which the valence of the graph is not fixed could be accommodated within the GFT formalism, by using several fields.}. Also the gauge invariance property fits in the same scheme. 

Imposing gauge invariance, we get
\begin{equation}
\Psi^{t}[\gamma, {\vec{g}}; 2t](\vec{h}) = \int (d\tilde{g})^{|V(\gamma)|}(dq)^{|E(\gamma)|} \prod_{v\in V(\gamma)} \Phi_{v} (\{ h_{v,i}\tilde{g}_{v}^{-1}g_{v,i}^{-1}q_{e_{v,i}}^{-1}\})
\end{equation}
implying, in turn,
\begin{equation}
\Psi^{t}[\gamma, {\vec{g}}; 2t](\vec{h}) = \int (dq)^{|E(\gamma)|} \prod_{v\in V(\gamma)} \phi_{v} (\{g_{v,i}^{-1}q_{e_{v,i}}^{-1}h_{v,i}\}), 
\end{equation}
where we have defined
\begin{equation}
\phi_{v}(\{ h_{i} \}) = \int d\tilde{g}_{v} \Phi({\tilde{g}_{v}h_{i}}), 
\end{equation}

This analysis shows that the group averaged coherent states associated to a graph might be written in terms of a group field theory, with a suitable number of fields according to the valence properties of the graph itself. Therefore, we will use these particular field configurations within GFT as a form of test configurations. However, before doing this, we need to go one more step further in the definition of the semiclassical LQG states.

\

In considering the coherent states for quantum gravity, in fact, one has to analytically continue the $g_{a}$ elements from $\SU$ to $\SLC$ in order to peak the states on configuration with given parallel transports and given triads per link. This analytic continuation exists and is unique \cite{Hall, coherentLQG}.

In  the case of complexified heat kernels (see appendix), the heat kernel itself is peaked around a $\SLC$ group variable. From here on, group elements denoted with a capital letter will correspond to $\SLC$ elements. The correct analytic continuation of the element associated to the link gives
$G_{ab} = G_{b}^{-1} G_{a}$.
In fact, using the isomorphism between $\SLC$ and $\mathcal{T}^*\SU$, the cotangent bundle of $\SU$, this $\SLC$ element is identified with a point in the phase space of LQG (or $\SU$ BF theory) at that spacetime point, or, in the discrete setting, associated to any given element of the discrete spacetime lattice. In particular, considering again the single link, $h_{ab} \in \SU$, while $g_{ab}$ is replaced by $G_{ab} \in \SLC$. 

In order to discuss the implications of the decomposition, we need to introduce some notation. We recall the polar decomposition of the $\SLC$ matrices (making the correspondence with $\mathcal{T}^*\SU$ explicit):
\begin{equation}
G_{ab} = g_{ab} \exp(E_{ab}), \qquad G_{a} = g_{a} \exp(E_{a}), \qquad G_{b} = g_{b} \exp(E_{b}),
\end{equation}
where $g_{X}$ are $\SU$ elements while $E_{X} = E_{X}^{i} \sigma_{i}$ are $\mathfrak{su}(2)$ matrices, with $\sigma_{i}$ denoting the Pauli matrices.

In the language of LQG, these $\SLC$ elements have a direct geometrical interpretation in terms of expectation values of holonomies and fluxes. More precisely, the $\SU$ part will correspond to the expectation value of the holonomy on the link, while the $E^{i}$ will be interpreted as the corresponding classical value for the triad/flux operator associated to an elementary surface dual to the link and intersecting it at a single point.

Using the above polar decomposition, we have
\begin{equation}
g_{ab}\exp(E_{ab}) =  \exp(-E_{b})g_{b}^{-1}g_{a} \exp(E_{a}) = g_{b}^{-1}g_{a} \exp(-\tilde{E}_{b}) \exp(E_{a}),
\end{equation}
where we have defined
\begin{equation}
\tilde{E}_{b} = (g_{b}^{-1}g_{a} )^{-1}E_{b} (g_{b}^{-1}g_{a}).
\end{equation}

{ Let us discuss briefly the geometrical interpretation of the above group elements, considering two oriented semi-links coming out of two vertices being glued to form an entire link. }

\begin{figure}
\begin{tikzpicture}[scale=2, >=stealth]
\draw[->, dashed] (0,0) .. controls (1,0.5) and (2,1.5) .. (3,1.5);
\filldraw (0,0) [black] circle (1pt);
\filldraw (3.02,1.5) [black] circle (1pt);
\path (0,0.1) node {$a$};
\path (3,1.6) node {$b$};
\draw [thick] (1.3,1) -- (1.6,0.8);
\draw [thick] (1.25,0.95) -- (1.55,0.75);
\filldraw (1.412,0.88) [gray] circle (1pt);
\draw[->] (0,0) .. controls (0.7,0.45) .. (1.35, 0.87);
\draw[->] (3,1.5) .. controls (2.7,1.45) and (2,1.25) .. (1.47, 0.89);
\path (1.1,0.2) node {$G_{a}=g_{a} \exp(E_{a})$};
\path (2.8,1.2) node {$G_{b}=g_{b}\exp(E_{b})$};
\path (0.1,0.3) node {$E_{a}$};
\path (2.8,1.7) node {$E_{b}$};
\path (1.1,1.) node {$E_{c}$};
\path (1.4,1.2) node {$E_{d}$};

\draw[->] (0.2,0.4) .. controls (0.7,0.8) .. (1., 1);

\draw[->] (2.7,1.7) .. controls (1.7,1.3) .. (1.5, 1.2);

\path (1.5,1) node {$P$};

\end{tikzpicture}
\caption{Gluing vertices}
\label{midpoints}
\end{figure}
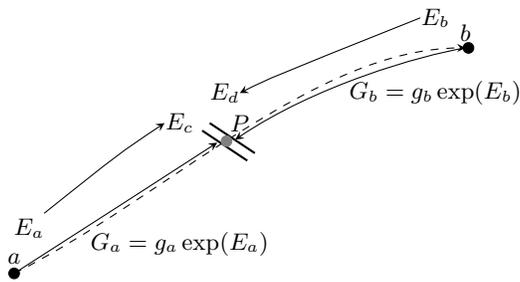

{ If we follow the geometrical interpretation according to which the Lie algebra elements $E_{a},E_{b}$ are (smeared) triads as seen from the vertices $a,b$, respectively, it is clear that they must be compared after appropriate transport on the midpoint at which the gluing takes place. Therefore, define (see figure \ref{midpoints}):
\begin{equation}
E_{c} = g_{a} E_{a} g_{a}^{-1}, \qquad E_{d} = g_{b} E_{b} g_{b}^{-1} 
\end{equation}
In absence of gauge invariance, we should expect that, in order to have a sensible geometrical interpretation, these two triads must coincide modulo a sign
(due to opposite orientations of the surfaces of integration),
\begin{equation}
E_{c} = -E_{d}.
\end{equation}
Taking the exponential of this equation we obtain:
\begin{equation}
g_{a} \exp(E_{a}) g_{a} ^{-1} = g_{b} \exp(-E_{b}) g_{b}^{-1},
\end{equation}
which leads to:
\begin{equation}
G_{a} g_{a}^{-1} = g_{b} G_{b}^{-1}.
\end{equation}
This last equation can be seen as a ``geometricity constraint'' on the data assigned to the vertices: if these constraints are satisfied, the links obtained by gluing the corresponding legs will be characterized by a very simple geometrical content, easily visualized in terms of gluing of cells (simplices, hypercubes or other kind of basic ``chunks of spacetime'') along common boundaries. 
This implies that:
\begin{equation}
E_{a} = - g_{a}^{-1} g_{b} E_{b} g_{b}^{-1} g_{a} = - \tilde{E}_{b}
\end{equation}
whence:
\begin{equation}
g_{ab}\exp(E_{ab}) = g_{b}^{-1}g_{a} \exp( 2 E_{a}),
\end{equation}
where the RHS is already a polar decomposition of the $\SLC$ element.
Therefore the link, after gluing, is peaked on the parallel transport obtained by correctly composing the elementary parallel transports, while the triads will be uniquely determined by the triad as seen from one vertex, as we would expect\footnote{It is interesting to note that, before imposing gauge invariance, we could be tempted to use these geometricity constraints to put more conditions on the data being assigned to the vertex functions. In particular, defining:
\begin{equation}
Q_{X} = G_{X} g_{X}^{-1},
\end{equation}
the geometricity constraints become:
\begin{equation}
Q_{a}= Q_{b}^{-1}
\end{equation}
whenever we want to glue vertex $a$ with vertex $b$ through a particular leg. If we want to keep the geometric interpretation independently from the particular gluing pattern, \ie if we want to keep the possibility of gluing every leg with any other leg, we obtain a rather strong constraint. It is enough to consider these constraints in cycles:
\begin{equation}
\left. \begin{array}{c} Q_{a} = Q_{b}^{-1} \\
Q_{b}^{-1} = Q_{c} \\
Q_{c} = Q_{a}^{-1} 
\end{array}
\right \}  \Rightarrow Q_{a} = Q_{a}^{-1}
\end{equation}
that in turn implies:
\begin{equation}
Q_{a} = \pm \mathbb{I} \Leftrightarrow E_{a} = 0,
\end{equation}
with the parallel transport left free. These constraints, then, would suggest that the only way in which the gluings are consistent with a naive geometrical intuition independently of the gluing patter is that {\it all the triads are vanishing}.  In fact, this is not the only possibility. One obvious way out is to include a
constraint on the pattern used for the gluings, in order to avoid the ciclicity that we have exploited to derive our conclusion. 
}.

Let us summarize briefly where we stand so far.
The semiclassical canonical wavefunctions associated to a given graph have been written in terms of products of functions associated to the vertices. In turn, these vertex functions will have a dependence on certain group variables associated to the vertex itself and to the each leg attached to it. The combinatorics is such that, when assembled, the total wavefunction depends only on certain combinations of these variables, matching the original dependence on group variables.

We have also seen that the possibilities for the peaks of the vertex functions are not always compatible with geometrical requirements. Rather, we
have to impose some constraints, which will select, among all the possible assignments of $\SLC$ elements, the ones compatible with a consistent gluing of elementary geometric cells along common boundaries.

Additionally, notice that whatever the Lie algebra elements are, even if the vertex wavefunctions are peaked on $\SU$ elements which are different from the identity, but they are the same link per link, the wavefunction in terms of the graph is peaked on the flat connection. Similarly, after complexification and consider heat kernels peaked on $\SLC$ elements, whenever they are equal, the wavefunction associated to the link is peaked on the identity of $\SLC$ itself, \ie not only on the flat connection, but also on zero triads. 

\subsection{Candidate GFT mean field configuration}

Having decomposed the coherent wave function associated to a generic graph in terms of vertex wave functions, we can now re-interpret this special vertex wave function as a classical GFT field, and use it in that context.

We will consider a compact Lie group $G$, and the group field theory defined from:
\begin{equation}
\phi : G^{m} \rightarrow \mathbb{C}
\end{equation}
and a classical action $S[\phi]$ or equivalently classical field equations:
\begin{equation}
\phi(\{h_{i}\}_{i=1}^{m}) + \sum_{w} \lambda_{w} \frac{\delta \mathcal{O}_{w}[\phi]}{\delta \phi(\{h_{i}\}_{i=1}^{m})}=0
\end{equation}
where $w$ is a label that identifies the various terms $\mathcal{O}_w$ in the action. We limit our analysis to the case in which $G = \SU$, for simplicity. We focus on the particular field configuration:
\begin{equation}
\xi_{\vec{g}}(\{h\};t) =  c(\lambda)\int dh \prod_{i=1}^{m} \rho_{t}(h_{i}g^{-1}_{i} h)
\end{equation}
where $g_{i}$ are group elements (in $\SU$ or in $\SLC$) on which the heat kernels would be peaked if the integration over the group variable associated to the vertex would be absent; with such gauge invariance restriction being imposed, the function will be peaked instead on gauge invariant equivalence classes of the $m$ group elements associated to the $m$ links of the vertex. We ask the following questions:
is $\xi$ a classical solution for some GFT? Is it an approximate solution for one of the cases we are interested in? What consequences, i.e. what restrictions on the parameters characterizing $\xi$, follow from demanding that it is in fact an approximate solution? In the corresponding regime of parameters, what is the effective dynamics for perturbations around such mean field configuration?
In the rest of the paper we try to address these questions.

{ To be more specific, we will consider several  possibilities for the mean field $\xi$, in dependence of the possible choices that we have in realizing the GFT theory in terms of field content (with colors or not) and gluings (specific gluing patterns or not). These choices are summarized as follows:
\begin{enumerate}
\item single field GFT \begin{description}
\item{a.} all the links are peaked on the same $\SLC$ element;
\item{b.} each link peaks on a generic $\SLC$ element;
\end{description}
\item colored GFT \begin{description}
\item{a.} for each color, all the links are peaked on the same $\SLC$ element;
\item{b.} for each color, each link peaks on a generic  $\SLC$ element.
\end{description}
\end{enumerate}

While the situation in case (1a.) leads necessarily to graph wavefunctions that are peaked on the identity in $\SLC$ for each complete link (i.e., after gluing), the others do not. According to the choice of group elements (which will be ultimately decided by the dynamics) and of the gluings (kinematical requirements on the particular GFT action), the outcome
might be a graph wavefunction which peaks, on each link, on nontrivial $\SLC$ elements.}

{Our (longer term) goal in fact is a bit more ambitious than this. We want to establish a pattern (closely related to the self consistent mean field method that we have discussed in the previous sections) which applies to the problem of finding the physical content of certain group field theories, in a suitable macroscopic (hydrodynamic) approximation, and to relate it to possible dynamics of effective geometries. We will try to understand what is the form of the dynamical equations determining the mean field values (\ie the peaks of the wavefunctions) once we make the educated {\it ansatz} that we should look for solutions in the form of heat kernel coherent states.
As in the case of many body systems, the equations of motion for the full field will provide an effective equation for the mean values/peaks. These peaks will play the role of {\it order parameters}: not only they will determine the properties of the geometry on which the state is peaked, but also they will determine the shape and the {\it symmetries} of the theory obtained by expanding the GFT around the heat kernels obtained in this way.
}

\section{$2$-dimensional case}
We start from a simple case, a GFT generating $2D$ simplicial complexes in its perturbative expansion, and corresponding to a quantization of topological BF theory with $\SU$ as gauge group. Despite its simplicity, it will represent our case study, on which we will try to model higher dimensional cases. 

Before attacking the GFT as such, it is worth recalling some basic facts of the classical theory behind it (see \cite{BF} for more details), that could be useful to give meaning to the later manipulations and results. The general form of the BF action is
\begin{equation}
S \, = \, \int_{\mathcal{M}} \Tr\left({B \wedge F(\omega)} \right) \label{BF}
\end{equation}
where $\mathcal{M}$ represents the spacetime D-manifold, $B$ is a $(D-2)$-form valued in the $\mathfrak{su}(2)$ Lie algebra, $\omega$ is a 1-form connection, also valued in the Lie algebra of the gauge group, and $F(\omega)$ is the corresponding 2-form curvature; the trace is taken in the fundamental representation of the algebra. The (spatial components of the) $B$ field and the (spatial components of the) connection $\omega$ are canonically conjugate variables. At the discrete level, with the connection 1-form replaced by group elements representing its holonomies, and $B$ field replaces by Lie algebra elements, one has that the classical phase space variables on each discrete element of the lattice is given by the cotangent bundle of $\SU$, expressed in terms of the same variables. The classical equations of motion are 
\begin{equation}
d_\omega(B) =0 \quad F(\omega) = 0
\end{equation}
enforcing the compatibility between connection $\omega$ and $B$ field, and flatness of the connection. It is then clear that the classical theory is trivial, as it contains only flat configurations. At the quantum level, the $B$ field acts as a Lagrange multiplier enforcing the same flatness condition, and the theory (for closed manifolds) amounts to an evaluation of the volume of the moduli space of flat connection, on the give topology $\mathcal{M}$.

The GFT framework (for the class of models we are going to consider) encodes the same quantum dynamics, by assigning amplitudes to any Feynman diagram dual to a given D-dimensional simplicial complex corresponding to the discrete path integral for the discrete version of the above action \cite{ioaristide}; on top of this, it embeds this dynamics in a wider framework which includes degrees of freedom coming from topology change (and singular complexes). No direct relation between the GFT equations of motion and the classical (continuum) BF equations of motion has been spelled out yet, though.

\subsection{Single-field GFT}

To begin with, let us consider what happens in the simplest case of a single field (non-colored GFT).

The basic field is a function of a pair of group variables, $\phi(g_{1},g_{2})$, with $g\in\SU$. 
The action we take is the one generating triangulated surfaces in the Feynman expansion for the partition function:
\begin{equation}
S= \frac{1}{2} \int dg_{a}dg_{b} \phi_{ab} \phi_{ba} + \frac{\lambda}{3} \int dg_{a}dg_{b}dg_{c} \phi_{ab}\phi_{ca}\phi_{bc},
\end{equation}
where we use the notation $\phi_{ab} = \phi(g_{a},g_{b})$ in order to avoid clutter. Note that we do not assume that $\phi_{ab}=\phi_{ba}$
The equation of motion obtained upon variation of this action is:
\begin{equation} \label{eom2d}
\phi(h_{2},h_{1}) +  \lambda\int dh_{3}  \phi(h_{2},h_{3}) \phi(h_{3},h_{1})=0.
\end{equation}
Notice also that if we had considered the interaction term containing, for example, the pairings $\phi_{ab} \phi_{ac}\phi_{bc}$, we would have obtained a different equation of motion. For a more detailed discussion of this point, see appendix \ref{appendixGFT}. 
For the present discussion, we have chosen a pairing that ensures orientability of the simplicial complexes obtained in the GFT Feynman expansion.
We consider now the heat kernel (coherent) states, in the 2D case, corresponding in the GFT context to bivalent vertices. Let us consider, as a trial configuration,
\begin{equation} \label{trial2d}
\xi_{G_a,G_b}(g_{1},g_{2};t) = \mu \int dh \rho_{t}(g_{1}h G_{a}) \rho_{t}(g_{2}h G_{b}),
\end{equation}
which is obtained by the convolution of two heat kernels (with the same spread) peaked on different elements of the (complexified) Lie group, $G_{a}^{-1}$ and $G_{b}^{-1}$ (thus, in our case $\SLC$ elements). Here $\mu$ is a normalization parameter to be fixed by the equations of motion.
Using the properties of the heat kernel,
\begin{equation} 
\xi(g_{1},g_{2}; t) = \mu \rho_{2t} (g_{1}g_{2}^{-1}G_{b}^{-1}G_{a}).
\end{equation}
This latter is in fact a heat kernel on the coset $\SU\times \SU/D_{2} \sim \SU$, where $D_{2} = \{ (h,h)  | h \in \SU \} \subset \SU \times \SU$ (for a complete discussion of heat kernels on coset spaces, see \cite{camporesi}).

It is clear that, if the two group elements $G_{a},G_{b}$ are the same, the field will be given by the heat kernel peaked around the identity in $\SLC$. Furthermore, it is obvious from this preliminary discussion that we will not be able to fix both $G_{a}$ and $G_{b}$: gauge invariance is telling us that one of them is redundant, i.e. the state will depend only on the equivalence class of elements under gauge transformations, labelled by $G_aG_b^{-1}$.
However, we will be able to fix the value of the product $G_{b}^{-1}G_{a}$ using the GFT equations of motion. 

We now have to plug the trial solution \eqref{trial2d} into the equation of motion \eqref{eom2d}. For simplicity in the notation, we denote the product $G_{b}^{-1}G_{a} = G$.
We get:
\begin{equation}
\mu \left[ \rho_{2t} (g_{2}g_{1}^{-1} G) + \lambda \mu \rho_{4t}(g_{2}g_{1}^{-1} G^{2}) \right] = 0.
\end{equation}
Clearly, $\mu=0$ is a solution to this equation. However, this particular solution represents the ``trivial'' GFT vacuum associated to the absence of any geometry, and, actually, of any space at all; it is the trivial \lq\lq Fock\rq\rq GFT vacuum. We are interested, then, in the vanishing of the other term in the equation of motion. To analyze its content, define $g=g_{1}g_{2}^{-1}$:
\begin{equation}
\rho_{2t}(gG) + \lambda \mu \rho_{4t}(gG^{2}) = 0 \quad \forall g \in \SU.
\end{equation}

Being valid for any group element $g$, this equation corresponds actually to an infinite tower of equations (one for each value of $g$) for the classical phase space variables $G$, with parameters $t$, $\lambda$ and $\mu$.  {\it These equations are then our \lq\lq geometrodynamics\rq\rq equations (in the simple BF case)}, obtained here as equations for the order parameters characterizing the GFT (mean field) background configuration in our (hydrodynamic) approximation, i.e. neglecting the contribution coming from fluctuations around the mean field, and coming directly from the full GFT dynamics.

\

We are not able, at the present stage, to recast the above equations into a more geometrically transparent form, or to make more direct contact with the classical BF theory equations. However, we can try to solve them, and hope to make contact with classical BF theory at least at the level of the solutions we find. 

It is immediate to realize that an exact solution to this equation is $t=0, \mu = -1/\lambda, G = \mathbb{I}$, \ie 
\begin{equation}
\xi_{\mathbb{I},\mathbb{I}}(g_{1},g_{2};0)= -\frac{1}{\lambda} \delta(g_{1}g_{2}^{-1}).
\end{equation}
Despite its obvious relevance, this solution is not really admissible because it corresponds to an infinite total action (and thus lies outside the space of fields for which the GFT model is defined). Therefore, we keep $t\neq 0$ and find some optimal value for the other free parameters of the heat kernel such that it is a reasonable approximation for a solution.
In order to do so, we expand the above equation in representations, using the Peter-Weyl decomposition of the heat kernel:
\begin{equation}
\rho_{t}(g) = \sum_{j} d_{j} e^{-t C_{j}} \chi_{j}(g), \qquad \chi_{j}(g) = t^{j}_{mn}(g) \delta_{mn}, \qquad C_{j}=j(j+1).
\end{equation}
Using this decomposition in the equation of motion, we get a tower of equations (labeled by $j$) for $t$ and $G$:
\begin{equation}
t^{j}_{nm}(G)+ \lambda \mu e^{-2t C_{j}} t^{j}_{nm}(G^{2})=0,
\end{equation}
or, equivalently,
\begin{equation}
\delta_{nm}+ \lambda \mu e^{-2t C_{j}} t^{j}_{nm}(G)=0.
\end{equation}
It is immediate to realize that the only $\SLC$ elements whose representation matrices are proportional to the identity for all representations $j$ are those belonging to its center, namely $\mathbb{I}$ and $-\mathbb{I}$. Indeed, their representation matrices are:
\begin{equation}
t^{j}_{nm} (\pm \mathbb{I}) = (\pm 1)^{2j} \delta_{nm}.
\end{equation}
The equation for $j=0$ provides the condition $\mu=-1/\lambda$, whence:
\begin{equation}
1- e^{-2t C_{j}} (\pm1)^{2j} = 0, \qquad j >0.
\end{equation}
Obviously, the exponential is always positive and smaller than one (since $t C_{j}>0$). Clearly, if $G=-\mathbb{I}$, when specializing to the case of half integer $j$,  the LHS of this equation would be always of order one, therefore failing to approach zero for any representation. Therefore, we have to keep $G=\mathbb{I}$ as the only plausible option.

However, unless $t=0$, the GFT equations cannot be satisfied. In particular, for fixed $t$, for representations $j$ is such that $t \lambda_{j} \geq 1$
we will have that the LHS of the equations are of order one. Therefore, our heat kernel with finite spread will represent an approximate solution to the equations of motion, as long as the spread is kept very small (mimicking, then, the exact solution given by the Dirac delta). The fact that this function is not a solution will be evident examining representations of sufficiently high spin, the threshold been given by the condition $2t \lambda_{j} \approx 1$. One has also to keep in mind that a spread of order $t$ in the $g$ (connection) variables implies a spread of order $1/t$ in the conjugate $B$ variables.

With these considerations in mind, we can reach some partial conclusions. The trial field profiles \eqref{trial2d}, with $t\approx 0$ and $G=\mathbb{I}$, can be seen as approximate solutions to the GFT field equation. Recall that $G=\mathbb{I}$ means $G_a=G_b$ in terms of the original GFT variables, which still implies that any wave function associated to a generic graph will have $G=\mathbb{I}$ as the phase space point associated to any graph link.  They represent  physical configurations of the (discrete) BF fields, that are sharply peaked around the trivial connection (everywhere on the graph), and peaked with large fluctuations a degenerate classical configuration of the $B$ field, $B=0$ everywhere. 
Once more, we cannot make yet explicit contact between our effective \lq\lq geometrodynamics\rq\rq and classical BF theory, but one should notice that this selected configuration is one of the classical solutions of BF theory: a trivial connection, up to gauge transformations, and any $B$ field satisfying $B^i(x)B_i(x) = const$ (the remaining components in the Lie algebra are pure gauge). So the candidate GFT configuration corresponding to LQG coherent states is a solution of the classical GFT dynamics if it corresponds to coherent states peaked on a degenerate (in the $B$), flat (with respect to the connection $\omega$) configuration\footnote{Recall the $B$ field plays the role of a Lagrange multiplier in both continuum and discrete formulations, so that large fluctuations around its classical value do not imply dynamical instabilities.}. Note also, once more, that, despite corresponding to a somewhat degenerate classical configuration, it does corresponds to a highly non-trivial quantum state, receiving contributions from arbitrary spin or group excitations, and in principle defined for arbitrarily complicated graphs, thus very far from the GFT free (Fock) vacuum. Still the result could be interpreted as suggesting that the full classical dynamics should be looked for in the dynamics of GFT perturbations around this degenerate man field configuration, rather than in the mean field equations themselves.

\subsection{Colored GFT} 

The next case, in order of complexity, is represented by the colored version of the same model\footnote{The addition of colors implies several constraints on the combinatorics of the resulting Feynman diagrams \cite{Razvan, Razvan2}, and is important from the point of view of GFT symmetries \cite{GFTdiffeo}, but it corresponds to the same Feynman amplitudes and thus is expected to correspond to the same classical BF theory.}, in which we have three fields, labelled by $R,G,B$, such that the action reads:
\begin{equation}
S=\frac{1}{2} \int dg_{a}dg_{b} \left( \phi_{ab}^{R} \phi_{ba}^{R}+\phi_{ab}^{G}\phi_{ba}^{G}+\phi_{ab}^{B}\phi_{ba}^{B} \right) + \lambda \int (dg)^{3} \phi_{ab}^{R} \phi_{bc}^{G} \phi_{ca}^{B}.
\end{equation}
The equations of motion are:
\begin{equation}
\phi^{R}_{ba} + \lambda \int \phi^{G}_{bc} \phi^{B}_{ca} dg_{c}=0;
\qquad
\phi^{G}_{ba} + \lambda \int \phi^{B}_{bc} \phi^{R}_{ca} dg_{c}=0;
\qquad
\phi^{B}_{ba} + \lambda \int \phi^{R}_{bc} \phi^{G}_{ca} dg_{c}=0.
\end{equation}
As in the previous case, it is immediate to check that:
\begin{equation}
\phi^{R}_{ab}=\frac{\sigma_{R}}{\lambda} \delta(g_{a}g_{b}^{-1}G_{R}); \qquad \phi^{G}_{ab}=\frac{\sigma_{G}}{\lambda} \delta(g_{a}g_{b}^{-1}G_{G}); 
\qquad
\phi^{B}_{ab}=\frac{\sigma_{B}}{\lambda} \delta(g_{a}g_{b}^{-1}G_{B})
\end{equation}
is an exact solution, provided that $\sigma_{R}\sigma_{G}\sigma_{B} = -1$ and three conditions for the parameters, namely $G_{R}=G_{G}G_{B}=\mathbb{I}$, and cyclic permutations. Using these conditions, it is easy to obtain
\begin{equation}
G_{R} = \zeta_{R} \mathbb{I}, \qquad G_{G} = \zeta_{G} \mathbb{I}, \qquad G_{B} = \zeta_{B} \mathbb{I},
\end{equation}
where $\zeta_{R},\zeta_{G},\zeta_{B}$ are signs satisfying the condition $\zeta_{R}\zeta_{G}\zeta_{B}=1$. However, once more one cannot consider the configurations as acceptable as such, because they lead to a divergent GFT action.

As in the case of a noncolored model, we can then consider the regularized case of heat kernels with a finite spread:
\begin{equation}
\phi^{R}_{ab} = \xi^{R}_{ab} = \mu_{R} \rho_{t}(g_{a}g_{b}^{-1} G_{R}),
\end{equation}
and similar expressions for the other colors. One gets:
\begin{equation}
\mu_{R} \rho_{t} (g_{b}g_{a}^{-1}G_{R}) + \lambda \mu_{G}\mu_{B} \int \rho_{t}(g_{b}g_{c}^{-1}G_{G})\rho_{t}(g_{c}g_{a}^{-1}G_{B}) dg_{c}=0
\end{equation}
After straightforward manipulations, this equation becomes:
\begin{equation}
\mu_{R} \rho_{t} (gG_{R}) + \lambda \mu_{G} \mu_{B} \rho_{2t}(g G_{B}G_{G})=0,
\end{equation}
while the other equations can be obtained by even permutations of $R,G,B$ labels. In terms of representations, we obtain the following three towers of equations:
\begin{equation}
\mu_{R} \delta_{mn}+ \lambda \mu_{G} \mu_{B} e^{-t \lambda_{j}} t^{j}_{mn}(G_{B}G_{G}G_{R}^{-1})=0,
\end{equation}
\begin{equation}
\mu_{G} \delta_{mn}+ \lambda \mu_{B} \mu_{R} e^{-t \lambda_{j}} t^{j}_{mn}(G_{R}G_{B}G_{G}^{-1})=0,
\end{equation}
\begin{equation}
\mu_{B} \delta_{mn}+ \lambda \mu_{R} \mu_{G} e^{-t \lambda_{j}} t^{j}_{mn}(G_{G}G_{R}G_{B}^{-1})=0.
\end{equation}

Once more, these are our effective \lq\lq geometrodynamics\rq\rq equations for the order parameters $G_X$ labelling classical phase space points (of BF theory), obtained from the mean field approximation of the fundamental GFT dynamics.

Again, taking first $j=0$ one gets:
\begin{equation}
\mu_{R} = \frac{\sigma_{R}}{\lambda}; \qquad \mu_{G} = \frac{\sigma_{G}}{\lambda}; \qquad \mu_{B} = \frac{\sigma_{B}}{\lambda}; \qquad  \sigma_{R}\sigma_{G}\sigma_{B}=-1
\end{equation}
Following the same reasoning of the scalar case, we get:
\begin{equation}
G_{B}G_{G}G_{R}^{-1} = \mathbb{I} \qquad
G_{R}G_{B}G_{G}^{-1} = \mathbb{I} \qquad
G_{G}G_{R}G_{B}^{-1} = \mathbb{I},
\end{equation}
which are the same that we have obtained in the case of the Dirac delta. Therefore, the structure of the solutions to the equations is the same, as far as the group elements associated to the peaks are concerned. 

The same comments on the trial configuration made in the case of the single-field model apply to this more complicated case. These heat kernels are not exact solutions, even though they can approach with arbitrary accuracy ($t \rightarrow 0$) an (unphysical) exact solution (but keep in mind the spread $1/t$ in the conjugate $B$ variable). 
The geometrical interpretation of the (unique) solution found is similar, so we do not repeat it, with the only exception that we now might have that some of the parallel transports are peaked on the $\SLC$ matrix $-\mathbb{I}$, which represents a full rotation of $2\pi$: in the case of half-integer representations, it leads to an overall multiplication by $-1$.

\section{3-dimensional GFT}

We can extend the previous analysis to any higher dimensional case. In particular, here we focus on the 3-dimensional case, namely, for the non-colored case, the Boulatov GFT model for 3d BF (1st order gravity), in the Euclidean signature \cite{boulatov}.
The Feynman amplitudes of this model are simplicial path integrals for the discrete 3d version of the action \eqref{BF}, or equivalently correspond to the Ponzano-Regge spin foam model \cite{ioaristide, iogft, review, alex}, in terms of a discrete triad and a discrete gravity connection. At the canonical level, the phase space is parametrized by such discrete triad, an $\mathfrak{su}(2)$ element, and the holonomy of the connection, an $\SU$ element, for each edge of the graph by means of which any spin network state is defined (see \cite{carlo, thomas} as well as \cite{fluxes}).

\subsection{Single-field case}

To begin with, we consider the simple (non-colored) model with a single field. The field will depend on three group elements, projected using the diagonal action of the group
\begin{equation}
\phi(g_{1},g_{2},g_{3}) = \int_{\SU} dh \, \Phi(g_{1}h ,g_{2} h,g_{3} h),
\end{equation}
in order to impose gauge invariance (closure of the triangle edge vectors in the metric representation).

The action is such that, in a perturbative expansion of a path integral, the Feynman diagrams are (dual to) oriented simplicial decompositions of three dimensional manifolds of arbitrary topology, and is given by:
\begin{equation}\label{boulatov}
S= \frac{1}{2} \int (dg)^{3} \phi_{123} \phi_{321} + \frac{\lambda}{4} \int (dg)^{6} \phi_{123} \phi_{156} \phi_{453} \phi_{426},
\end{equation}
where the obvious notation has been employed. The reader is referred to appendix
\ref{appendixGFT} for a careful discussion of the possible alternative definitions of the model, for what concerns the combinatorics of field arguments. Our particular choice \eqref{boulatov}, among those ensuring orientability, leads to the simplest equations of motion, reducing the number of terms to be manipulated.

As in the 2d case, we will try to find (possibly, approximate) solutions in the form of the selected trial functions, dependent on a certain number of geometric parameters. The (highly nonlocal) field equation for the GFT will be then turned into equations for the geometric parameters, providing thus a sort of {\it geometrodynamics}.

In our case, the trial function is the convolution of heat kernels:
\begin{equation}
\xi_{abc} = \mu \int dh \rho_{t}(g_{a}h G_{a}) \rho_{t}(g_{b}hG_{b}) \rho_{t}(g_{c}hG_{c}),
\end{equation}
where $G_{a},G_{b},G_{c}$ are $\SLC$ elements encoding the geometrical properties of the mean configuration on which the state $\xi$ is peaked. In particular, in this three dimensional case, the relation between these group elements and the classical phase space of three dimensional geometry is transparent. In a polar decomposition of the $\SLC$ matrices, the $\SU$ part will be associated to the mean value of the parallel transport around which the state fluctuates, while the positive Hermitian part will be associated to the average triads/fluxes.

After simple manipulations, the equation of motion can be put into the form\footnote{It is worth mentioning that the expression in the interaction term would change dramatically if we were to change the interaction term in the original action, for instance by replacing $\phi_{123}$ by $\phi_{231}$ (an even permutation of the group variables, preserving the orientation properties). First of all, the equation of motion changes, involving more terms that have different combinatorial structures. Second, as a consequence of this first point, the various convolutions of heat kernels will be such that some (possibly all) of the heat kernels of the form $\rho_{2t}(h_{i}^{-1}h_{j})$ will be replaced by $\rho_{2t}(h_{i}^{-1}h_{j} G_{\alpha})$, where $G_{\alpha} = G_{i'}^{-1}G_{j'}$.}
:
\begin{equation*}
\mu \left[ \int dh  \rho_{t}(g_{3}h G_{a}) \rho_{t}(g_{2}hG_{b}) \rho_{t}(g_{1}hG_{c})  +
\right.
\end{equation*}
\begin{equation} 
+\left. \lambda \mu^{2} \int (dh^{3})  \rho_{t}(g_{1}h_{1} G_{a}) \rho_{t}(g_{2}h_{2}G_{b}) \rho_{t}(g_{3}h_{3}G_{c}) \rho_{2t}(h_{1}^{-1}h_{2})\rho_{2t}(h_{2}^{-1}h_{3})\rho_{2t}(h_{3}^{-1}h_{1})\right]=0
\end{equation}

Coming back to our concrete problem, as in the 2d case the choice $\mu=0$ gives an exact solution, associated to the ``trivial'' GFT vacuum. In order to find other solutions, we decompose the equation into a tower of equations by expanding it into representations. The result is:
\begin{equation}
\threej{j_{1}}{j_{2}}{j_{3}}{n_{1}}{n_{2}}{n_{3}} \threej{j_{1}}{j_{2}}{j_{3}}{r_{1}}{r_{2}}{r_{3}} t^{j_{1}}_{r_{1}m_{1}}(G_{c})t^{j_{2}}_{r_{2}m_{2}}(G_{b}) t^{j_{3}}_{r_{3}m_{3}}(G_{a}) + \lambda \mu^{2} \Omega_{n_{1}n_{2}n_{3}r_{1}r_{2}r_{3}} t^{j_{1}}_{r_{1}m_{1}}(G_{a})t^{j_{2}}_{r_{2}m_{2}}(G_{b}) t^{j_{3}}_{r_{3}m_{3}}(G_{c})=0,
\end{equation}
where the Wigner's $3j$ symbols have been introduced\footnote{It is worth noticing the crucial way in which the particular choices for the various terms in the action affect the way in which the $G_{a},G_{b},G_{c}$ enter this equation.
If we had chosen a kinetic term with the pairing $\phi_{123}\phi_{123}$, we would make this equation totally insensitive to the $\SLC$ elements.
In the case we are considering, the equation is insensitive to $G_{b}$ only. We do not have a clear understanding of this feature of the GFT dynamics.}.
To simplify the form of the equation, we have defined the tensor
\begin{equation}
\Omega_{n_{1}n_{2}n_{3}r_{1}r_{2}r_{3}} = \int (dh^{3})  t^{j_{1}}_{n_{1}r_{1}}(h_{1})t^{j_{2}}_{n_{2}r_{2}}(h_{2})t^{j_{3}}_{n_{3}r_{3}}(h_{3}) \rho_{2t}(h_{1}^{-1}h_{2})\rho_{2t}(h_{2}^{-1}h_{3})\rho_{2t}(h_{3}^{-1}h_{1}).
\end{equation}
This tensor can be re-written as:
\begin{equation}
\Omega_{n_{1}n_{2}n_{3}r_{1}r_{2}r_{3}} = \left[ \sum_{456} d_{j_{4}}d_{j_{5}}d_{j_{6}} e^{-2t (\lambda_{4}+\lambda_{5}+\lambda_{6})}
\sixj{j_{1}}{j_{2}}{j_{3}}{j_{4}}{j_{5}}{j_{6}}^{2}
 \right] \threej{j_{1}}{j_{2}}{j_{3}}{n_{1}}{n_{2}}{n_{3}} \threej{j_{1}}{j_{2}}{j_{3}}{r_{1}}{r_{2}}{r_{3}},
\end{equation}
where the $6j$ symbols have been introduced.

One of the $3j$ symbols factorizes away, leaving us with
\begin{equation}
\threej{j_{1}}{j_{2}}{j_{3}}{r_{1}}{r_{2}}{r_{3}} + \lambda \mu^{2} f(j_{1},j_{2},j_{3};t) t^{j_{1}}_{s_{1}r_{1}}(G_{a}G_{c}^{-1}) \delta_{s_{2}r_{2}} t^{j_{3}}_{s_{3}r_{3}}(G_{c}G_{a}^{-1}) \threej{j_{1}}{j_{2}}{j_{3}}{s_{1}}{s_{2}}{s_{3}} = 0 \quad \forall j_1,j_2,j_3.
\end{equation}
Here, for convenience, we have defined:
\begin{equation}
f(j_{1},j_{2},j_{3};t) = \sum_{456} d_{j_{4}}d_{j_{5}}d_{j_{6}} e^{-2t (\lambda_{4}+\lambda_{5}+\lambda_{6})}
\sixj{j_{1}}{j_{2}}{j_{3}}{j_{4}}{j_{5}}{j_{6}}^{2}
\end{equation}

These are the {\it geometrodynamics equations} for the order parameters derived from the GFT dynamics. Once more, we are not able yet to relate these equations directly with the classical BF equations, nor to give them a more geometrically explicit form. We then turn to the problem of identifying some solutions, which could instead be given a geometric interpretation.

In order for the equation of motion to be satisfied, we need:
\begin{equation}
t^{j}_{mn}(G_{a}G_{c}^{-1}) = c^{j}_{m} (G_{a}G_{c}^{-1}) \delta_{mn} \quad \forall j,
\end{equation}
\ie that the matrices are diagonal for any $j$. The properties of the representations we are playing with (see appendix \ref{appendixreps}) lead then to the conclusion:
\begin{equation}
G_{a}G_{c}^{-1} = \left( \begin{array}{cc} \alpha & 0 \\ 0 & 1/\alpha \end{array} \right), \qquad \alpha \in \mathbb{C}\setminus{0}
\end{equation}
Using this in the equation of motion, we can get rid of the second $3j$ symbol to get:
\begin{equation}
1 + \lambda \mu^{2} f(j_{1},j_{2},j_{3};t) c^{j_{1}}_{m_{1}}(G_{a}G_{c}^{-1}) c^{j_{3}}_{m_{3}}(G_{a}G_{c}^{-1}) = 0 \forall j_1,j_2,j_3, m_1, m_3 \, . 
\end{equation}
Therefore, we can argue that $c^{j_{1}}_{m_{1}} (G_{a}G_{c}^{-1})= c^{j_{1}}_{m_{1}'}(G_{a}G_{c}^{-1})$, \ie that the matrices are not only diagonal, but proportional to the identity matrix, in any representation. As we have already stated, the only $\SLC$ elements whose representation matrices are proportional to the identity in every representations are $\pm \mathbb{I}$.

Consider $j_{1},j_{2},j_{3} = 0$. Then one gets:
\begin{equation}
\mu^{2} = -\frac{1}{\lambda},
\end{equation}
which fixes the normalization of the field profile.

From its expression, it is manifest that, if $t\in \mathbb{R}$, $f>0$. Therefore, we get
\begin{equation}
1- f(j_{1},j_{2},j_{3};t) (\pm 1)^{2j_{1}} (\pm 1)^{2j_{3}}=0,
\end{equation}
that can be approximately be satisfied if the second term on the LHS is always negative, for all the representations. We get then that $G_{a}G_{c}^{-1} = \mathbb{I}$, thus fixing the geometric part that can be controlled by the equation of motion (EOM).

Therefore we obtain that the coherent state GFT configuration is a solution of the GFT dynamics for $G_a = G_c$ and for arbitrary $G_b$, or equivalently, taking into account the gauge invariance properties of the field, we can conclude that the solutions are parametrized by a single $\SLC$ element $G=G_bG_a^{-1}$. The dynamics is therefore clearly much richer than in the 2d case already examined, as one would expect. The case $G=\mathbb{I}$ would correspond once more to a degenerate 3d geometry, but we now see that other solutions are allowed. One should now attempt to prove that the only freedom is in the phase space variables corresponding to the discrete triad, while the above configurations all correspond to flat connections, as one would expect from classical BF theory. We leave this for further investigations.
Finally, conditions on $t$ must be obtained by the analysis of the closeness of $f$ to 1, similarly to what we have done in the case of 2D models. This requires a careful examination of the properties of the function $f$, that we also leave for future work. 

\subsection{Colored model}
The investigations about the symmetries of GFTs \cite{GFTdiffeo} have shown that, in order to implement within a GFT model a symmetry that corresponds to 3D simplicial diffeomorphisms, the colored generalization of the GFT model \cite{Razvan} is necessary, in which one has a multiplet of dynamical fields, each labeled by a color index. The quantum dynamics still generates random orientable 3D complexes of arbitrary topology, and it has been shown that the coloring also leads to the absence many singular configurations with respect to the usual models \cite{Razvan, Razvan2, vincentcolored}; this combinatorial properties of the resulting Feynman diagrams were in fact the original motivation for introducing coloring in the GFT framework. 

We consider the simplest incarnation of the theory, defined by the following action:
\begin{equation}
S= \frac{1}{2} \int (dg)^{3} \left[ \phi^{R}_{123}\phi^{R}_{321}+\phi^{G}_{123}\phi^{G}_{321}+\phi^{B}_{123}\phi^{B}_{321}+\phi^{V}_{123}\phi^{V}_{321} \right]+
\lambda \int (dg)^{6} \phi^{R}_{123}\phi^{G}_{156}\phi^{B}_{426}\phi^{V}_{543}.
\end{equation}
The analysis of the equations of motion goes exactly as in the case of the single-field model. However, there are some crucial differences that make this model much more complicated, and the extraction of geometrical information significantly more involved. 

Concerning the trial functions, as in the 2d case, the $\SLC$ peaks are different color by color, at least in principle:
\begin{equation}
\xi_{abc}^{X} = \mu_{X} \int dh \rho_{t}(g_{a}h G_{a}^{X})\rho_{t}(g_{b}h G_{b}^{X})\rho_{t}(g_{c}h G_{c}^{X}),
\end{equation}
where $X=R,G,B,V$. This implies that the structure of the tensor $\Omega$ that we have introduced in the previous section gets completely modified. Besides possessing an obvious color index, it depends in a non-trivial way on various products of $\SLC$ elements. For instance, the equation for the color $R$ reads:
\begin{equation*}
\mu_{R} \int dh \rho_{t}(g_{3}h G_{a}^{R})\rho_{t}(g_{2}h G_{b}^{R})\rho_{t}(g_{a}h G_{c}^{R}) + \lambda\mu_{G} \mu_{B} \mu_{V} \int (dh)^{3}\left[
\rho_{t}(g_{1}h_{1}G_{a}^{G})\rho_{t}(g_{2}h_{2}G_{b}^{B}) \rho_{t}(g_{3}h_{3}G_{c}^{V}) \times \right.
\end{equation*}
\begin{equation}
\times
\left.
\rho_{2t}(h_{2}^{-1}h_{3}G_{a}^{V}(G_{a}^{B})^{-1})
\rho_{2t}(h_{3}^{-1}h_{1}G_{b}^{G}(G_{b}^{V})^{-1})
\rho_{2t}(h_{1}^{-1}h_{2}G_{c}^{B}(G_{c}^{G})^{-1}) \right]=0
\end{equation}
In a decomposition in representations analogous to the one used to treat the scalar case, there are two obvious differences. First of all, the multiplication of the equations by the representation matrices of the elements $(G_{i}^{R})^{-1}$ does not eliminate any of the matrices contracted with the analogous of the
tensor $\Omega$. Furthermore, the latter has a dependence on the $\SLC$ elements such that it cannot be easily rewritten in terms of $3j$ and $6j$ symbols. These complications prevent us, at the moment, from identifying the solutions of the equations. We can only stress the fact that the classical dynamics for the order parameters we have obtained seems richer than in the single-field case.

However, it is easy to see that if we impose the constraint $G_{i}^{X}=G_{i}^{X'}$, then the analysis of the solutions goes exactly in the same way of the single-field case, producing exactly the same results.

The only difference involves the different normalizations. It is immediate to realize that: $\mu_{X} = \sigma_{X} \mu$, with $\sigma_{X}$ some signs, and
$\mu=|\lambda|^{-1/2}$. The signs are constrained to satisfy $\sigma_{R}\sigma_{G}\sigma_{B}\sigma_{V}=-\mathrm{sign}(\lambda)$.

\section{Effective dynamics around the heat kernel configuration}
{ In the previous sections we have considered some aspects of the mean field approximation applied to GFTs, at leading order, i.e. ignoring fluctuations around the mean field configurations. In order to understand its limits, and to start characterizing the full dynamics of the system around these non-perturbative vacua, it is important to elucidate the features of the effective theory for the small deviations away from the mean field. This is also needed to understand the stability of the configurations, and hence their viability as (nonperturbative) ``ground states'' for the GFT. Furthermore, in light of the possibility of expressing any GFT quantum dynamics in spin foam representation (which amounts to the perturbative expansion of the same in Feynman amplitudes written as functions of group representations), this analysis should clarify some of the features of the spin foam models that one would obtain around the new vacua and the differences with the ``fundamental'' ones. 
In the following, we will consider such effective dynamics in the same cases in which we have performed the analysis of the equations of motion and obtained conditions for the heat kernel GFT configurations. However, it should be clear, by now, that the analysis can be generalized to higher dimensional cases without any conceptual difficulty.}

\subsection{2D - single-field model}
{ The easiest case is the two dimensional simple GFT (without colors).
We will use the splitting of the field into a mean field part (the heat kernel satisfying the mean field conditions $\mu=-1/\lambda, G=\mathbb{I}$) and a fluctuation $\varphi$ (satisfying the same diagonal gauge invariance as the original field):
\begin{equation}
\phi_{ab} = \xi(g_{a},g_{b}) + \varphi(g_{a},g_{b}), \qquad \xi(g_{a},g_{b}) =- \frac{1}{\lambda} \rho_{2t} (g_{a}g_{b}^{-1}).
\end{equation}
The action for the field generates an effective action for $\varphi_{ab}$ which has the simple form:
\begin{equation*}
S_{\mathrm{\eff}}[\varphi]:= S[\xi+\varphi] - S[\xi] =
\end{equation*}
\begin{equation}
= \frac{1}{2} \int dg_{a} dg_{b} (\varphi_{ab})^{2} + \int dg_{a}dg_{b} \xi_{ab} \varphi_{ab} + \lambda  \int dg_{a} dg_{b} dg_{c}\left\{
\varphi_{ab}\xi_{bc} \xi_{ca} + \varphi_{ab} \varphi_{bc} \xi_{ca} + \frac{1}{3} \varphi_{ab}\varphi_{bc} \varphi_{ca}
  \right\}.
\end{equation}
In this effective action, we have a term, linear in $\varphi_{ab}$, which would vanish if $\xi_{ab}$ were a solution. However, as we have seen, $\xi_{ab}$
is not exactly a solution and hence it will give a (possibly small) effective contribution to the equations of motion. By taking care of the factors of $\lambda$, it is easy to see that these terms are of order $1/\lambda$: in these sense they are nonperturbative. The value of $t$ must be tuned in such a way that this term becomes negligible. We will consider this as done.

The kinetic term is completed by a quadratic term in $\varphi_{ab}$ which is obtained by convolution of the interaction term of the original theory with a single copy of the background solution $\xi_{ab}$. This means, and this is a first crucial and generic feature of the effective theories we obtain, that this term brings the full microscopic dynamics of the underlying GFT model (as well as the properties of the particular background solution considered) into the effective dynamics of the perturbations, already at the level of the effective kinetic term
\begin{equation}
\frac{1}{2}\int \varphi_{ab} \varphi_{cd} \mathcal{K}^{abcd} (dg)^{4} \quad . \end{equation} 

The latter has the explicit form:
\begin{equation}
\mathcal{K}^{abcd} = \delta(g_{a}g_{c}^{-1})\delta(g_{b}g_{d}^{-1}) - 2  \delta(g_{b}g_{c}^{-1})\rho_{2t}(g_{a}^{-1}g_{d}) \, =\,  \delta(g_{a}g_{c}^{-1})\delta(g_{b}g_{d}^{-1}) - 2  \int dg_e dg_f \mathcal{V}(g_a,g_b,g_c,g_d,g_e,g_f)\,\rho_{2t}(g_{e}^{-1}g_{f}).
\end{equation} 
Here, the disappearance of $\lambda$ is somehow fictitious. In fact, the conditions imposed on the parameters of the heat kernels (our effective \lq\lq geometrodynamics\rq\rq equations) are tying together $G$, $t$ and $\lambda$.
Let us finally note that,
in the case of two dimensional theories, the effective interaction term is just the same as the one present for the full theory, and hence its simplicial interpretation, even if not obviously useful or appropriate (given that now we may have already generated an effective spacetime by means of the choice of nonperturbative vacuum) is unchanged. 

It is useful to rewrite the action in terms of group representations, using the properties of the representation functions of $\SU$ (see appendix \ref{appendixreps}, or, for a more complete treatment, \cite{vilenkin}), and the consequent expression:
\begin{equation}
\varphi_{ab} = \sum_{j} Y^{j}_{mn} t^{j}_{mn}(g_{a}g_{b}^{-1}).
\end{equation}
The effective action reads:
\begin{equation}
S= \frac{1}{2} \sum_{j} \left[ \frac{1}{d_{j}} Y^{j}_{mn}{Y}^{j}_{nm}   -\frac{2}{d_{j}} e^{-2t C_{j}} Y^{j}_{mn}{Y}^{j}_{nm}   + \frac{\lambda}{3d_{j}} Y^{j}_{mn}
Y^{j}_{nr}Y^{j}_{rm} \right] \; ,  \quad C_j = j (j + 1) \quad .
\end{equation}
We have an effective dynamics with a new, non-trivial propagator. In this simple case, this is the only formal modification with respect to the fundamental (microscopic) GFT action. Obviously, this will result in modified spin foam amplitudes. The nature of the microscopic GFT model as a tower of infinite unitary matrix models with dimension $N = 2j +1$ is retained in the effective dynamics.

The diagonal nature of the kinetic term allows to re-write the same action in terms of a trivial one, by performing the rescaling:
\begin{equation}
Y^{j}_{mn}\rightarrow Z^{j}_{mn} = (1-2e^{-2t C_{j}})^{1/2} Y^{j}_{mn}  \quad .
\end{equation}
The resulting action can be considered as coinciding with the microscopic one, but in terms of a new, renormalized (or better, momentum dependent) coupling constant:
\begin{equation}
\lambda \rightarrow \lambda_{\eff}(j)= (1-2e^{-2t C_{j}})^{-3/2} \lambda
\end{equation}
The rescaled matrices can be seen as the components in representation space of a  ``quasiparticle'' field:
\begin{equation}
\omega(g) = \sum_{j} Z^{j}_{mn}t^{j}_{nm}(g),
\end{equation}
related to the original fluctuation field by a nonlocal transformation:
\begin{equation}
\omega(g) = \int T(g,h)\varphi(h) dh, \qquad T(g,h)= \sum_{j} d_{j} (1-2e^{-2t C_{j}})^{1/2}  \chi_{j}(gh^{-1}).
\end{equation}

The amplitudes of the effective theory can be immediately derived by making use of the Feynman rules for the corresponding matrix model (see \cite{mm}). In particular, the amplitude associated with a 2D simplicial complex $\Gamma$, dual to a fat graph obtained from the spin $j$ matrix model will be:
\begin{equation}
A(\Gamma,j) = \frac{1}{S} \lambda_{\eff} ^{F} d_{j}^{\chi(\Gamma)} = \frac{1}{S} \lambda^{F} (1-2e^{-2t C_{j}})^{-3F/2}d_{j}^{\chi(\Gamma)},
\end{equation}
where $F$ is the number of faces (triangles) of the complex, $\chi(\Gamma)$ is its Euler characteristic, and $S$ is a symmetry factor.

\paragraph*{Stability} The analysis of the scalar case, with its extreme simplicity, allows us to partially address the issue of the stability of the heat kernel solutions. In particular, it appears that there is an instability for low spins, given that
\begin{equation}
1-2 e^{-2t C_{j}} \leq 0, 
\end{equation}
showing that the coefficients in front of the kinetic term for such low spins is negative. Notice that this is even more the case the more one approaches the regime $t \approx 0$, which is the one in which $\xi$ is indeed a solution of the GFT equations.
This shows that the heat kernel GFT configurations can be used as good approximations of exact solutions, but they are probably perturbatively unstable: the stability matrix has negative eigenvalues. If this is the case, it would be interesting to study in more detail the physics behind this instability, as it may signal the breakdown of the GFT hydrodynamic approximation, and indirectly give insights on the physical meaning of the same.

\subsection{2D - colored model}

In the case of the colored model, the presence of several fields complicates the situation slightly. Indeed, the microscopic GFT interaction term induces a quadratic term characterized by a mixing between colors. 

The effective action written in group representations reads:
\begin{equation*}
S_{\eff} =  \sum_{j} \frac{1}{2d_{j}} \Big{[} R^{j}_{mn}{R}^{j}_{nm} +G^{j}_{mn}{G}^{j}_{nm}+B^{j}_{mn}{B}^{j}_{nm} + 
\end{equation*}
\begin{equation}
+ 2 e^{-2t \lambda_{j}} \left( \sigma_{B} G^{j}_{mn}R^{j}_{nr} t^{j}_{rm}(G_{B}) +  \sigma_{R} B^{j}_{mn}G^{j}_{nr} t^{j}_{rm}(G_{R}) +
 \sigma_{G} R^{j}_{mn}B^{j}_{nr} t^{j}_{rm}(G_{G}) \right) +
\frac{2\lambda}{d_{j}} G^{j}_{mn}R^{j}_{nr}B^{j}_{rm}
\Big{]}
\end{equation}
It is a tower (labeled by the irreducible representations of $\SU$) of three-matrix models with a standard interaction term, accompanied by a number of terms (in which the background field $\xi$ enters explicitly) generating oscillations between the colors. These terms introduce a form of nontrivial dynamics even if we truncate the effective action to the lowest order in $\lambda$. It is worth stressing that these effects are directly induced by the GFT interaction term due to the presence of a nontrivial background. In the simplicial, discrete spacetime interpretation of the resulting perturbative Feynman amplitudes (once more, not necessarily appropriate in this context) these terms is that they generate oscillations of 1d simplicial spaces (or bivalent spin networks) in coloring, while keeping the topology fixed. 

\

The diagonalization of this oscillating kinetic term can, however, be performed, at least in the case $G_{R}=G_{B}=G_{G}=\mathbb{I}$ (which is the case in which the heat kernel GFT configuration is a solution of the mean field dynamics), once it is recognized that the quadratic term can be rewritten as:
\begin{equation}
(R^{j},G^{j},B^{j}) \left( \begin{array}{ccc}
1 & \sigma_{B} \beta_{j} & \sigma_{G} \beta_{j} \\
\sigma_{B} \beta_{j} & 1 & \sigma_{R}\beta_{j}\\
\sigma_{G} \beta_{j} & \sigma_{R} \beta_{j} & 1
\end{array} \right) \left( \begin{array}{c}R^{j} \\ G^{j} \\ B^{j} \end{array}\right),
\label{kineticterm}
\end{equation}
where $\beta_{j}=e^{-2t C_{j}}$.

The possibility, if not need, for such diagonalization in field space is the only new feature of the colored model with respect to the single-field GFT.

One has to introduce new matrices, which are linear combinations of the representation matrices $R^{j},G^{j},B^{j}$ (the mapping being an orthogonal transformation, since the matrix in the kinetic term \eqref{kineticterm} is real and symmetric). In terms of these new matrices, the effective action will have a trivial kinetic term, at the price of complicating the cubic interaction term.
The transformation needed to put the kinetic term in canonical form depends on the representations, through the coefficient $\beta_{j}$.
Notice that for $j \rightarrow \infty$, the transformation reduces to the identy. This is due to the fact that for large $j$ the off-diagonal terms are exponentially suppressed. Therefore, the largest effect is on small representations, the crossover scale $J$ being determined by the condition $2 t C_{J} \approx 1$.

The fact that the transformation of the matrices are $j$-dependent has important implications in determining the propagating modes of the effective model.
Let us be schematic. We started from some fields:
\begin{equation}
\varphi^{I}_{ab} = \sum_{j} X^{I,j}_{mn} t^{j}_{nm}(g_{a}g_{b}^{-1}), \qquad X^{I,j}_{mn} = d_{j} \int \varphi^{I}_{ab} \overline{t^{j}_{mn}}(g_{a}g^{-1}_{b}),
\end{equation}
and we have discovered that the normal (truly propagating) modes are not given in terms of $X^{I}$ but of
\begin{equation}
Y^{\alpha, j}_{mn} = M^{\alpha I}_{j} X^{I,j}_{mn},
\end{equation}
where $M^{\alpha I}$ are the matrices diagonalizing the kinetic term \eqref{kineticterm}.
The matrices $Y$s can be then associated to fields:
\begin{equation}
\omega^{\alpha}(g) = \sum_{j,n,m} Y^{\alpha,j}_{mn} t^{j}_{nm}(g),
 = \int dh \, \varphi^{I}(h) \, K^{\alpha I}(h,g),
\end{equation}
a nonlocal linear relation between the two kinds of fields, controlled by the kernel
\begin{equation}
K^{\alpha I}(h,g) = \sum_{j,n,m} d_{j} M^{\alpha, I}_{j}  \overline{t^{j}_{mn}}(h) t^{j}_{nm}(g),
\end{equation} 

This is very similar to the relation between quasiparticles and atoms in the case of condensed matter systems. The effective quanta associated to the oscillations around the given (geometrical) GFT background are clearly collective modes of the fundamental GFT quanta.

\subsection{3D model(s)}

The discussion of the effective action for the three dimensional case goes along the same lines of the two dimensional case. The effective action is given by:
\begin{equation*}
S_{\eff}[\varphi] \equiv S[\xi+\varphi]-S[\xi] = \left \langle  \frac{\delta S}{\delta \phi_{abc}}  \Big|_{\phi = \xi} \varphi_{abc} \right \rangle + \frac{1}{2}
\left \langle  \frac{\delta^{2} S}{\delta \phi_{abc} \delta\phi_{def}}  \Big|_{\phi = \xi} \varphi_{abc} \varphi_{def} \right \rangle +
\end{equation*}
\begin{equation}
+
\frac{1}{3!} \left \langle  \frac{\delta^{3} S}{\delta \phi_{abc}\delta \phi_{def}\delta\phi_{ghi}}  \Big|_{\phi = \xi} \varphi_{abc} \varphi_{def} \varphi_{ghi} \right \rangle
+\frac{\lambda}{4} \left \langle \varphi_{abc}\varphi_{aef} \varphi_{dbf} \varphi_{dec} \right \rangle,
\end{equation}
where we are using a notation in which $\langle... \rangle$ denotes the integration with respect to the relevant group variables.

We have already mentioned the term linear in $\varphi$, which is not exactly zero unless $\xi$ is a solution of the equation of motion. However, by tuning $t$ we may control the effect of this term and make it negligible. Therefore the first nontrivial term in our expansion come from the second variation of the interaction term evaluated on $\xi$.

In the end, the effective action of the model (neglecting the linear term) reads:
\begin{equation}
S_{\eff} = \frac{1}{2} \int \varphi_{123} \varphi_{321} + \frac{\lambda}{2} \int \left( \varphi_{abc}\varphi_{aef} \xi_{dbf} \xi_{dec} +
\varphi_{abc} \varphi_{dbf} \xi_{aef} \xi_{dec} + 
\varphi_{abc} \varphi_{dec} \xi_{aef} \xi_{dbf} \right)+
\end{equation}
\begin{equation}
+\lambda \int (dg)^{6} \varphi_{123} \varphi_{156} \varphi_{426} \xi_{453} + \frac{\lambda}{4} \int (dg)^{6} \varphi_{123} \varphi_{156} \varphi_{426} \varphi_{453}.
\end{equation}

The main points to notice are the following, all concerning the new non trivial kinetic term:
 
1) due to the expansion of the theory around a nontrivial background configuration, the fundamental GFT interaction term percolates down to the quadratic and the cubic part of the effective action; also,  each insertion of background field leads to a multiplication by $|\lambda|^{-1/2}$, which means that the quadratic term is of order $\lambda^{0}$;  

2) the three new terms contributing to the kinetic term, in addition to the trivial one, are not equivalent: they correspond to three distinct combinatorial structures. Moreover, they corresponds to oscillations in representation labels along the lines of propagation. 

However, contrary to simpler 2D case, it is not yet clear whether it is possible to envisage a transformation, that would put the kinetic term in a canonical diagonal form.
The nature of the oscillations, as the Feynman amplitudes resulting from the effective action, can be studied in representation space. One has to use the conditions on $\xi$:
\begin{equation}
\lambda < 0; \qquad \xi_{abc} = \frac{1}{|\lambda|^{1/2}}  \int dh \rho_{t} (g_{a}h G_{a}) \rho_{t}(g_{b}hG_{b}) \rho_{t}(g_{c}hG_{a}),
\end{equation}
with $G_{a},G_{b}$ arbitrary elements of $\SLC$, and the expansion 
\begin{equation}
\varphi_{123} = \sum_{\{j\}} X^{j_{1}j_{2}j_{3}}_{n_{1}n_{2}n_{3}} \threej{j_{1}}{j_{2}}{j_{3}}{r_{1}}{r_{2}}{r_{3}} t^{j_{1}}_{r_{1}n_{1}}(g_{1})
t^{j_{2}}_{r_{2}n_{2}}(g_{2})t^{j_{3}}_{r_{3}n_{3}}(g_{3}),
\end{equation} 

Once more, the detailed analysis would not add much to the present discussion.

We therefore restrict our attention only to the mixing terms in the quadratic part. 
 Let us consider just one of them:
\begin{equation}
\lambda \int \varphi_{abc}\varphi_{aef} \xi_{dbf} \xi_{dec}  (dg)^{6} = \lambda \int \varphi_{abc}\overline{\varphi_{aef}} \xi_{dbf} \xi_{dec}  (dg)^{6} = (-1)  \times
\end{equation}
\begin{equation}\int dg_{a}
\sum X^{j_{1}j_{2}j_{3}}_{n_{1}n_{2}n_{3}} \threej{j_{1}}{j_{2}}{j_{3}}{r_{1}}{r_{2}}{r_{3}} 
 \overline{X^{j_{4}j_{5}j_{6}}_{n_{4}n_{5}n_{6}}} \threej{j_{4}}{j_{5}}{j_{6}}{r_{4}}{r_{5}}{r_{6}} 
t^{j_{1}}_{r_{1}n_{1}}(g_{a})\overline{ t^{j_{4}}_{r_{4}n_{4}}(g_{a}) }W^{j_{2}j_{3}j_{5}j_{6}}_{r_{2}r_{3}r_{5}r_{6},n_{2}n_{3}n_{5}n_{6}}.
\end{equation}
where we have introduced
\begin{equation*}
W^{j_{2}j_{3}j_{5}j_{6}}_{r_{2}r_{3}r_{5}r_{6},n_{2}n_{3}n_{5}n_{6}}=
\int (dg)^{5} (dh)^{2}
\left[
t^{j_{2}}_{r_{2}n_{2}}(g_{b}) t^{j_{3}}_{r_{3}n_{3}}(g_{c})
\overline{t^{j_{5}}_{r_{5}n_{5}}(g_{e}) t^{j_{6}}_{r_{6}n_{6}}(g_{f})} \times \right.
\end{equation*}
\begin{equation}
\times
\left.
\rho_{t}(g_{d}h_{1}G_{a})\rho_{t}(g_{e}h_{1}G_{b}) \rho_{t}(g_{c}h_{1}G_{c})
\rho_{t}(g_{d}h_{2}G_{a})\rho_{t}(g_{b}h_{2}G_{b}) \rho_{t}(g_{f}h_{2} G_{c}) \right],
\end{equation}
to (slightly) simplify the notation. We might want to use $G_{a}=G_{c}$, and gauge invariance to fix $G_{b}=\mathbb{I}$.
Orthonormality of the representations is expressed as:
\begin{equation}
\int dg_{a} t^{j_{1}}_{r_{1}n_{1}}(g_{a})\overline{ t^{j_{4}}_{r_{4}n_{4}}(g_{a}) } = \frac{1}{d_{j_{1}}} \delta^{j_{1}j_{4}} \delta_{r_{1}r_{4}} \delta_{n_{1}n_{4}}
\end{equation}
In turn, this implies that we can reduce the expression to:
\begin{equation}
\sum X^{j_{1}j_{2}j_{3}}_{n_{1}n_{2}n_{3}} \threej{j_{1}}{j_{2}}{j_{3}}{r_{1}}{r_{2}}{r_{3}} 
 \overline{X^{j_{1}j_{5}j_{6}}_{n_{1}n_{5}n_{6}}} \threej{j_{1}}{j_{5}}{j_{6}}{r_{1}}{r_{5}}{r_{6}} 
W^{j_{1};j_{2}j_{3}j_{5}j_{6}}_{r_{2}r_{3}r_{5}r_{6},n_{2}n_{3}n_{5}n_{6}},
\end{equation}
or
\begin{equation}
\sum X^{j_{1}j_{2}j_{3}}_{n_{1}n_{2}n_{3}} 
\overline{X^{j_{1}j_{5}j_{6}}_{n_{1}n_{5}n_{6}}} 
W^{j_{2}j_{3}j_{5}j_{6}}_{n_{2}n_{3}n_{5}n_{6}},
\end{equation}
where:
\begin{equation}
W^{j_{1};j_{2}j_{3}j_{5}j_{6}}_{1;n_{2}n_{3}n_{5}n_{6}} = 
\sum_{r_{1}}
\threej{j_{1}}{j_{2}}{j_{3}}{r_{1}}{r_{2}}{r_{3}} 
 \threej{j_{1}}{j_{5}}{j_{6}}{r_{1}}{r_{5}}{r_{6}} 
W^{j_{2}j_{3}j_{5}j_{6}}_{r_{2}r_{3}r_{5}r_{6},n_{2}n_{3}n_{5}n_{6}}.
\end{equation}
The structure for the other terms in the quadratic part of the effective action is the same, with differences only in the orderings/combinatorial structure. We could summarize the structure that we obtain into the expression:
\begin{equation}
\frac{1}{2} X^{j_{1}j_{2}j_{3}}_{n_{1}n_{2}n_{3}}\overline{X^{j_{4}j_{5}j_{6}}_{n_{4}n_{5}n_{6}}} \mathcal{K}^{j_{1}j_{2}j_{3}j_{4}j_{5}j_{6}}
_{n_{1}n_{2}n_{3}n_{4}n_{5}n_{6}}, 
\end{equation}
where
\begin{equation}
\mathcal{K}^{j_{1}j_{2}j_{3}j_{4}j_{5}j_{6}}
_{n_{1}n_{2}n_{3}n_{4}n_{5}n_{6}} =  
^{0}\mathcal{K}^{j_{1}j_{2}j_{3}j_{4}j_{5}j_{6}}
_{n_{1}n_{2}n_{3}n_{4}n_{5}n_{6}}
- 2 \left( \delta^{j_{1}j_{4}} \delta_{n_{1}n_{4}} W^{j_{1};j_{2}j_{3}j_{5}j_{6}}_{1;n_{2}n_{3}n_{5}n_{6}}
+
 \delta^{j_{2}j_{5}} \delta_{n_{2}n_{5}} W^{j_{2};j_{1}j_{3}j_{4}j_{6}}_{2;n_{1}n_{3}n_{4}n_{6}}
+
 \delta^{j_{3}j_{6}} \delta_{n_{3}n_{6}} W^{j_{3};j_{1}j_{2}j_{4}j_{5}}_{3;n_{1}n_{2}n_{4}n_{5}}
 \right)
\end{equation}
and where
\begin{equation}
^{0}\mathcal{K}^{j_{1}j_{2}j_{3}j_{4}j_{5}j_{6}}
_{n_{1}n_{2}n_{3}n_{4}n_{5}n_{6}}=
\delta^{j_{1}j_{6}} \delta_{n_{1}n_{6}} \delta^{j_{2}j_{5}} \delta_{n_{2}n_{5}} \delta^{j_{3}j_{4}}\delta_{n_{3}n_{4}}
\end{equation}
is just the original kinetic term.

The entire dependence on the structure of the interaction term and on the background solutions, \ie $t$ and the various group elements of $\SLC$, is encoded in the tensors $W_{1},W_{2},W_{3}$. While the explicit structure of the quadratic term is not helping much, still the processes it encodes are rather clear. Indeed, we could write down the Feynman diagrams of the emergent theory at once, starting from the structure of the Feynman diagrams of the fundamental theory, constructed out of the basic propagator and vertex in figure 4.

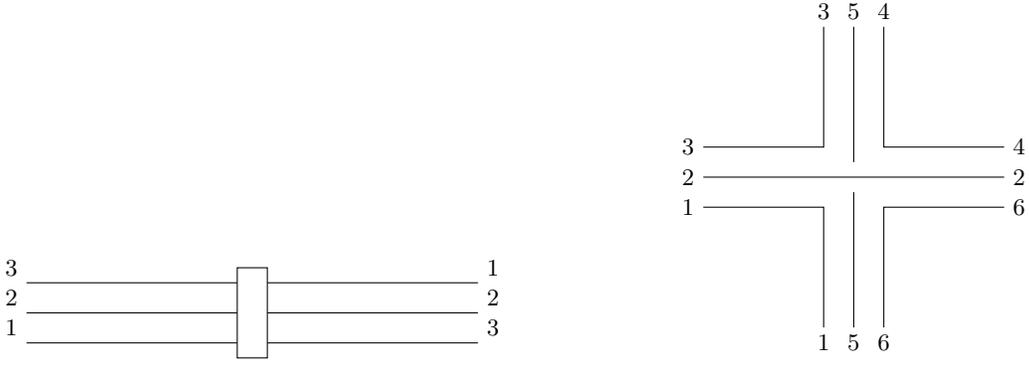
\begin{figure}
\begin{tikzpicture}[scale=2, >=stealth]
\draw (0,0) -- (1.4,0);
\draw (0,0.2) -- (1.4,0.2);
\draw (0,0.4) -- (1.4,0.4);
\draw (1.6,0) -- (3,0);
\draw (1.6,0.2) -- (3,0.2);
\draw (1.6,0.4) -- (3,0.4);
\draw (1.4,-0.1) rectangle (1.6,0.5);
\path (-0.1,0.1) node {$1$};
\path (-0.1,0.3) node {$2$};
\path (-0.1,0.5) node {$3$};
\path (3.1,0.1) node {$3$};
\path (3.1,0.3) node {$2$};
\path (3.1,0.5) node {$1$};
%\path (3,1.6) node {$b$};
%\path (1.3,1) node {$h_{ab}$};
%\path (1.4,0.7) node {$g_{ab}$};
\end{tikzpicture}
%\caption{The basic propagator.}
%\label{prop}
%\end{figure}
%\end{minipage}
\hspace{2cm}
%\begin{minipage}[r]{5cm}
%\begin{figure}
\begin{tikzpicture}[scale=2, >=stealth]
\draw (0,0.2)--(0.8,0.2)--(0.8,1);
\draw (0,0)--(2,0);
\draw (0,-0.2)--(0.8,-0.2)--(0.8,-1);
\draw (1,1) -- (1,0.1);
\draw (1,-0.1)--(1,-1); 
\draw (1.2, 1)-- (1.2, 0.2) -- (2,0.2);
\draw (1.2,-1)-- (1.2,-0.2) -- (2,-0.2);
\path (-0.1,0.2) node {$3$};
\path (-0.1,0) node {$2$};
\path (-0.1,-0.2) node {$1$};
\path (2.1,0.2) node {4};
\path (2.1,0) node {2};
\path (2.1,-0.2) node {6};
\path (0.8,1.1) node {3};
\path (1,1.1) node {5};
\path (1.2,1.1) node {4};
\path (0.8,-1.1) node {1};
\path (1,-1.1) node {5};
\path (1.2,-1.1) node {6};
\end{tikzpicture}
\caption{Basic propagator and vertex}
\label{vertex}
\end{figure}
%\end{minipage}

Apart from the symmetry factors, the various induced diagrams can be obtained very easily by heat kernels insertions on the external legs, in all the possible ways. Therefore, the three oscillatory contributions to the kinetic term will correspond to the diagrams in figure \ref{effpropdiagram}.

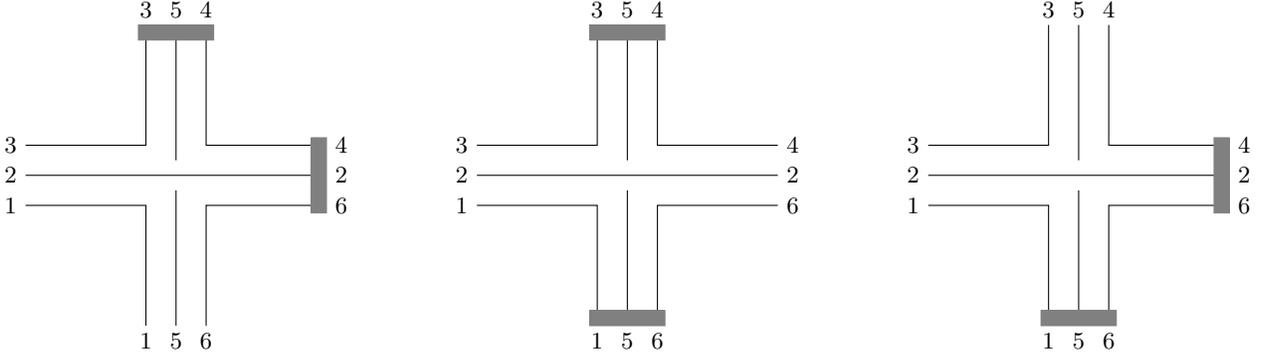
\begin{figure}
\begin{tikzpicture}[scale=2, >=stealth]
\draw (0,0.2)--(0.8,0.2)--(0.8,1);
\draw (0,0)--(2,0);
\draw (0,-0.2)--(0.8,-0.2)--(0.8,-1);
\draw (1,1) -- (1,0.1);
\draw (1,-0.1)--(1,-1); 
\draw (1.2, 1)-- (1.2, 0.2) -- (2,0.2);
\draw (1.2,-1)-- (1.2,-0.2) -- (2,-0.2);
\path (-0.1,0.2) node {$3$};
\path (-0.1,0) node {$2$};
\path (-0.1,-0.2) node {$1$};
\path (2.1,0.2) node {4};
\path (2.1,0) node {2};
\path (2.1,-0.2) node {6};
\path (0.8,1.1) node {3};
\path (1,1.1) node {5};
\path (1.2,1.1) node {4};
\path (0.8,-1.1) node {1};
\path (1,-1.1) node {5};
\path (1.2,-1.1) node {6};
\filldraw[gray] (0.75,0.9) rectangle (1.25,1);
\filldraw[gray] (1.9,0.25) rectangle (2,-0.25);
% Second diagram
\draw (3,0.2)--(3.8,0.2)--(3.8,1);
\draw (3,0)--(5,0);
\draw (3,-0.2)--(3.8,-0.2)--(3.8,-1);
\draw (4,1) -- (4,0.1);
\draw (4,-0.1)--(4,-1); 
\draw (4.2, 1)-- (4.2, 0.2) -- (5,0.2);
\draw (4.2,-1)-- (4.2,-0.2) -- (5,-0.2);
\path (2.9,0.2) node {$3$};
\path (2.9,0) node {$2$};
\path (2.9,-0.2) node {$1$};
\path (5.1,0.2) node {4};
\path (5.1,0) node {2};
\path (5.1,-0.2) node {6};
\path (3.8,1.1) node {3};
\path (4,1.1) node {5};
\path (4.2,1.1) node {4};
\path (3.8,-1.1) node {1};
\path (4,-1.1) node {5};
\path (4.2,-1.1) node {6};
\filldraw[gray] (3.75,0.9) rectangle (4.25,1);
\filldraw[gray] (3.75,-0.9) rectangle (4.25,-1);
% Third diagram
\draw (6,0.2)--(6.8,0.2)--(6.8,1);
\draw (6,0)--(8,0);
\draw (6,-0.2)--(6.8,-0.2)--(6.8,-1);
\draw (7,1) -- (7,0.1);
\draw (7,-0.1)--(7,-1); 
\draw (7.2, 1)-- (7.2, 0.2) -- (8,0.2);
\draw (7.2,-1)-- (7.2,-0.2) -- (8,-0.2);
\path (5.9,0.2) node {$3$};
\path (5.9,0) node {$2$};
\path (5.9,-0.2) node {$1$};
\path (8.1,0.2) node {4};
\path (8.1,0) node {2};
\path (8.1,-0.2) node {6};
\path (6.8,1.1) node {3};
\path (7,1.1) node {5};
\path (7.2,1.1) node {4};
\path (6.8,-1.1) node {1};
\path (7,-1.1) node {5};
\path (7.2,-1.1) node {6};
\filldraw[gray] (7.9,0.25) rectangle (8,-0.25);
\filldraw[gray] (6.75,-0.9) rectangle (7.25,-1);
\end{tikzpicture}
\caption{The three contributions to the kinetic term: the gray rectangles correspond to insertions of heat kernels. It is understood that a sum is performed over the lines that are beginning and ending at one heat kernel insertion.}
\label{effpropdiagram}
\end{figure}

Nonetheless, let us stress once more that these peculiar contributions to the kinetic term are encoding a nontrivial dynamics, not associated to any change in the structure of the graph depicting the combinatorial structure of the pairings of field arguments, but only affecting its associated representation labels. 

\

This very same analysis applies to the case of colored 3D GFTs. Without going into the explicit calculations (very similar to the scalar case), we can immediately say that one additional feature will have to be added. The percolation of the original interaction term (combining triangles with different colors into a tetrahedron) onto the effective kinetic term will result not only in an effective dynamics that at the lowest order (free theory) will change the representations labelling the lines of propagation, but also in oscillations in the colorings, in a way similar to the case of 2D colored models.
The quadratic part of the effective action reads:
\begin{equation*}
\frac{1}{2} \int (dg)^{3} \left[ \varphi^{R}_{abc}\varphi^{R}_{cba}+\varphi^{G}_{abc}\varphi^{G}_{cba}+\varphi^{B}_{abc}\varphi^{B}_{cba}+\varphi^{V}_{abc}\varphi^{V}_{cba} \right] + 
\end{equation*}
\begin{equation*}
+\lambda \int (dg)^{3} \left[ 
\varphi^{R}_{abc} \varphi^{G}_{aef} \xi^{B}_{dbf} \xi^{V}_{dec}+
\varphi^{R}_{abc} \xi^{G}_{aef} \varphi^{B}_{dbf} \xi^{V}_{dec}+
\varphi^{R}_{abc} \xi^{G}_{aef} \xi^{B}_{dbf} \varphi^{V}_{dec}+\right.
\end{equation*}
\begin{equation}
\left.
\xi^{R}_{abc} \varphi^{G}_{aef} \varphi^{B}_{dbf} \xi^{V}_{dec}+
\xi^{R}_{abc} \varphi^{G}_{aef} \xi^{B}_{dbf} \varphi^{V}_{dec}+
\xi^{R}_{abc} \xi^{G}_{aef} \varphi^{B}_{dbf} \varphi^{V}_{dec}  \right],
\end{equation}
where the $\xi^{X}$ are determined as in the previous section. The percolation of the interaction term of the fundamental theory into the effective quadratic term, and hence in the effective propagator is even more evident.

}

\section{Fundamental vs effective symmetries}

In any field theory, the choice of the vacuum is crucial for the understanding of the symmetries of the effective theory describing the dynamics of the excitations around that vacuum. Viceversa, the behaviour under symmetry transformation gives important informations about the physical nature of the chosen vacuum itself. In particular, the ground state defines an order parameter (or a set of order parameters) determining the symmetry group of the truncated theory. Nonetheless, it is important to remember that the effective theory for fluctuations around the new ground state remains invariant under the original symmetry group: the difference is in the {\it representation} of the symmetry group, which becomes nonlinear.

Let us clarify this point. Consider a field theory (e.g. a GFT) for some fields collectively denoted by $\Phi$ (internal/tensor/spinor indices are omitted). Assume that the theory is invariant under the action of a symmetry transformation:
\begin{equation}
\Phi \rightarrow \Phi' = U\Phi, \qquad S[U\Phi]=S[\Phi]
\end{equation}

The Fock vacuum $\Phi = 0$ is obviously invariant under the action of $U$.

Imagine that the new (non-perturbative) ground state is now \emph{not} invariant under the action of the symmetry, but, for example, only transforms covariantly:
\begin{equation}
U \Phi_{0} \neq \Phi_{0} \quad .
\end{equation}
When we consider the effective action that we obtain expanding around $\Phi_{0}$ 
\begin{equation}
\Phi=\Phi_{0}+ \phi
\end{equation} 
we obtain that, under the action of the symmetry group:
\begin{equation}
\Phi' = U \Phi \Rightarrow \phi' = \Phi'-\Phi_{0} = U \phi + (U-\mathbb{I}) \Phi_{0} \quad .
\end{equation}
Notice that here we have written exact expressions. We have never referred to the fact that $\phi$ is small, in any sense, or that $U$ is an infinitesimal transformation.
Then, the effective action for $\phi$ is not invariant under $\phi\rightarrow U \phi$, but rather it is invariant under the transformation 
\begin{equation}
\phi \rightarrow U\phi +
(U-\mathbb{I})\Phi_{0} \quad .
\end{equation}

This must be expected whenever an order parameter transforming under some representation of the symmetry group of the theory is introduced. 

In our GFT case, the heat kernels are parametrized by the group elements around which they are peaked. In turn, they determine, by the mechanism we have just outlined, the shape of the symmetries of the effective theory around the chosen vacuum.

Let us discuss further this important point, in the case of 3d colored GFT, where the implementation of (simplicial) diffeomorphism symmetry has been recently unraveled \cite{GFTdiffeo}. It has been shown that the GFT action is left invariant by global (from the QFT point of view) transformations (forming a quantum group) of the GFT field, which correspond (at the level of the Feynman amplitudes, given by simplicial gravity path integrals, to $3d$ simplicial diffeomorphisms \cite{simplicialDiffeos}: in terms of simplicial complexes appearing in the perturbative expansion, they correspond to independent translations of the vertices \cite{GFTdiffeo}. Their geometric meaning is manifest in the non-commutative metric representation of GFTs \cite{ioaristide}.

These transformations, in the group representation, take the form:

\begin{equation}
\phi(g_{1},g_{2},g_{3}) \rightarrow \exp[\Tr( \epsilon_{12}g_{1}g_{2}^{-1} ] \exp[\Tr(\epsilon_{23}g_{2}g_{3}^{-1} )]
\exp[\Tr(\epsilon_{31}g_{3}g_{1}^{-1} ) ]\phi(g_{1},g_{2},g_{3})   \label{diffeo}
\end{equation}

where $\epsilon_{ij}$ are $\mathfrak{su}(2)$ Lie algebra elements, and the trace is taken in the fundamental representation (the phases are given by the {\it non-commutative plane waves} of \cite{PR3}, coming in turn from the quantum group Fourier transform \cite{laurentmajid, karim}, used in \cite{ioaristide} to obtain the non-commutative metric representation of GFTs), and where we neglected the color labels, which, although crucial for the definition and understanding of the symmetry, are not relevant for our present exposition.

As we have mentioned, the presence of a nontrivial vacuum would change the action of these transformations. In particular, we could consider the action onto the heat kernel configurations we have studied in this paper.
%\subsection{Dirac}

\

Before studying the coherent GFT configuration, let us study the simpler case
\begin{equation}
\psi(g_{1},g_{2}g_{3}) = \int dh \delta(g_{1}hG_{a})\delta(g_{2}hG_{b})\delta(g_{3}hG_{c}) \quad .
\end{equation}

The transformed field reads:
\begin{equation}
\psi_{123}' = \exp(i\Tr[\epsilon_{12} G_{a}^{-1}G_{b}])\exp(i\Tr[\epsilon_{23} G_{b}^{-1}G_{c}])\exp(i\Tr[\epsilon_{31} G_{c}^{-1}G_{a}]) \psi_{123},
\end{equation}
which is equal to $\psi_{123}$ if and only if:
\begin{equation}
G_{a}=G_{b}=G_{c}
\end{equation}
Therefore, the field configuration obtained by means of a convolution of Dirac deltas, in the general case, transforms covariantly: it gets multiplied by a phase. However, the phase turns out to be just the identity if all the the elementary distributions used to construct the field are peaked the same group elements.

This is an instance of a more general fact. The group elements entering the field configuration (around which we expand the theory) as parameters are also influencing the properties under symmetry transformations, allowing, for instance, for different phases.  We can have three different phases:
\begin{description}
\item{I.} $G_{a}=G_{b}=G_{c}$, and the field configuration is invariant under the full group of transformations \eqref{diffeo}. This represent a diffeo-invariant solution.
\item{II.} Two of the group elements are equal, \eg $G_{a}=G_{b}$. Therefore, the symmetry group is broken down to a smaller subgroup (generated by the only $\epsilon_{12}$). We can interpret this case as a configuration possessing an isometry.
\item{III.} $G_{a} \neq G_{b} \neq G_{c}$, which corresponds to the completely anisotropic configuration, and to a purely covariant field configuration.
\end{description}

The above discussion, even if it refers to a field configuration which is unphysical, clarifies the logic that has to be followed in determining the symmetry properties of any particular GFT configuration one is interested in. In the case of heat kernels (which, we recall, become delta functions in the $t=0$ limit), it is immediate to see that, under such GFT diffeo transformations:

1) the heat kernel (coherent state) configuration is not left invariant, and transforms in the general way \eqref{diffeo} in which any GFT field transforms;

2) a a consequence, it is not transformed into another heat kernel peaked around the diffeo-transformed group element. This can be seen, for example, by taking the spread $t$ to zero, and checking that the function obtained in this way is \underline{not} a Dirac delta. 

Thus, on the one hand, as we expect from their interpretation as (second quantized) wave functions approximating a general continuum metric, the GFT coherent state configurations do not possess any special isometry, in the general case, and transform covariantly under GFT diffeos; the only case in which an invariance under diffeos is obtained corresponds to their $t=0$ limit, and to the special case $G_a=G_b=G_c$, which, as we had seen, can be interpreted as a solution of the classical GFT dynamics (somehow expected to encode also the projection onto diffeo invariant states, from the canonical point of view) corresponding, however, to degenerate geometries. On the other hand, the transformed configuration does not peak either on diffeo-transformed canonical phase space points, so it somehow does not seem reproduce faithfully the action of diffeomorphisms, one would expect from the LQG (canonical quantum gravity point of view). For instance, this is at odds with the transformation properties of the canonical LQG \lq\lq condensate\rq\rq representation studied in \cite{Tim, Hanno}.

\

Another interesting field configuration, similar to the above ones, that is worth discussing briefly is the {\it exact} solution of the GFT dynamics in 3d given by the field:
\begin{equation}
\psi(g_1,g_2,g_3) = \int \, dh \delta(g_1 h)\, f(g_2 h)\, \delta(g_3 h) \, = \, f( g_2 g_1^{-1}) \, \delta( g_3 g_1^{-1}) \quad , \quad \int | f |^2 = 1
\end{equation}
identified first in \cite{eterawinston}. It is easy to verify (see also \cite{EteraFloGFT}) that this field configuration is invariant under a subset of GFT diffeos \eqref{diffeo} (generated by $\epsilon_{31}$) which, together with the additional rotation invariance that can also be  identified at the GFT level \cite{GFTdiffeo,EteraFloGFT}, form a deformed Poincar\'e invariance corresponding to the Drinfeld double quantum group $D\SU$. This confirms the interpretation of the same GFT configuration as a {\it quantum flat space} used in \cite{eterawinston} for the interpretation of the effective field theory for (reduced) GFT perturbations as a matter field theory on a non-commutative flat space (see also the analogue construction in the 4d Lorentzian case in \cite{ioeteraflorian}, but confirms also the general scheme for classifying GFT phases in terms of their symmetry properties, that we outlined above.

\

The sketchy analysis that we have reported here shows concretely how
\begin{itemize}
\item the non-perturbative vacuum one could chose (on physical or mathematical grounds) as the mean field configuration relevant for the hydrodynamic/continuum approximation of the GFT dynamics (and as the tentative description of the \lq\lq quantum space condensate\rq\rq), will, in general, not be invariant under diffeomorphisms;
\item in the resulting effective theory, the action of diffeomorphisms is not the linear one but rather is of the nonlinear type we described;
\item the effective theories can be classified in phases, by means of the residual symmetry, if any, of the configuration around which we expand the theory. 
\end{itemize}

These facts should then be taken into account when trying to derive an effective dynamics of geometry from the fundamental GFT dynamics,  if we expect the effective theory to be given by (some modified form of) Einstein's General Relatvity, which is characterized by its symmetry under diffeomorphisms. 

On top of this, we have to add another feature that we have uncovered in the discussion of the effective theories for GFT perturbations. The symmetry that we are considering here is expressed in terms of the field $\varphi$ which is \emph{not diagonalising} the kinetic term of the effective field theory. The true effective symmetry should instead be identified from a combination of the effective symmetry group for the field $\varphi$ \underline{\it and} of the Bogoliubov-like transformation that diagonalizes the kinetic term (needed to get the true propagating modes), defining the physical field $\psi$.

\section*{Summary, discussion and outlook}
In this paper we have made a first tentative step towards the extraction of an effective classical geometrodynamics from group field theory hydrodynamics, using mean field theory techniques as applied to quantum GFT. 

\

This meant first of all identifying a candidate macroscopic GFT configuration with characteristic order parameters endowed with a geometric interpretation. This was chosen as the vertex building block of loop quantum gravity semi-classical coherent states associated to arbitrary graphs \cite{coherentLQG}. The coherent states constructed out of this, and associated to graphs with N vertices are then interpreted, in the analogy with condensates, as N-particle states.

We have then used this candidate GFT configuration as the mean field around which to expand in an hydrodynamic approximation of the microscopic GFT model, and obtained the relevant consistency hydrodynamic equations (the GFT analogue of the Gross-Pitaevski equations  adapted to this reference field). These, in turn, become equations for the geometric order parameters (identifying points on the classical phase space of gravity/BF theory) and other constants entering the definition of the background GFT field.
Although their relation with the classical spacetime theory behind the GFT models considered (2d and 3d BF theory) is not yet entirely clear, these equations can be interpreted as a form of classical geometrodynamics, here obtained directly from the GFT dynamics.

The special nature of heat kernel coherent states is crucial. Therefore, the problem arising at this point would be to find a dynamical mechanism that selects these states among all the possible ones. This task requires considerations that are beyond the limited scope of this paper, involving the detailed examination of the full path integral (and a better understanding of GFT perturbative renormalization \cite{GFTrenorm}).

\

The only solutions to these equations that we have been able to identify seem to correspond to degenerate geometries/B variables and flat connections, with quantum uncertainty on the latter being very small and on the former being correspondingly very large. As we pointed out in the text, this result is compatible with the classical dynamics of BF theory. However, three points have to be made in this respect: 1) the complexity of the equations, and of the corresponding solutions (thus the classical geometries selected) seem to grow considerably with the dimension of the spacetime they refer to, as we would expect; 2) the addition of \lq\lq coloring\rq\rq seem to allow for further geometric content, that is however not easy to elucidate; 3) the full geometric content of a many-particle quantum state constructed out of our reference wavefunction, contrary to the case of ordinary condensates, involves also a gluing operation, depending on the graph one wants to reconstruct, on our reference wavefunctions associated to spin network vertices; this plays an important role in the computation of geometric operators, but it plays no role in our construction, thus making the geometric interpretation of the identified configurations trickier.

It must also be stressed that even a background configuration peaked on degenerate geometries as the one we identified is a highly non-trivial state from the point of view of the fundamental theory. It corresponds to a possibly continuum space in which geometry is everywhere defined, even if degenerate, made out of a possibly infinite number of microscopic GFT quanta (spin network vertices), whose fundamental degrees of freedom are all excited (e.g. the states are obtained by a sum over arbitrary group representations). By way of contrast, the microscopic GFT vacuum is a no-space state composed of no GFT quanta at all, where all geometric operators are identically vanishing.

\

Despite their approximate nature, these solutions still allows to gain some better understanding of the theory. Indeed, the classical action itself (and hence the equations derived from it) can be seen as a saddle point/stationary phase approximation to the full path integral that defines the quantum gravity model. In the language of quantum effective actions, the classical action is only the lowest order contribution to the formal semiclassical expansion, and therefore even exact solutions to the classical equation of motion will have the role of approximate solutions for the full theory. 
Consequently, it is more significant to find general conditions according to which the desired function approximates reasonably well a solution, rather than an exact solution itself. 

This is true even in the case of BEC. The mean field approximation (MFA) gives a good qualitative description of the phenomena associated to Bose--Einstein condensation, but in certain conditions, it simply fails, due to instabilities appearing in systems with large inhomogeneities or large time derivatives \cite{castindum}. In other words, only for some kind of situations and for some kind of observables, a MFA is adequate in giving the correct semiquantitative results. 

In the case of GFT, we expect that, in the full quantum/statistical problem, the classical action will receive important corrections and hence that even exact solutions to the classical equation of motion will be able to capture only a portion of the physical features, and possibly only under certain conditions. In the next section, we will see explicit signals of the approximate nature of the theory in the appearance of some instabilities, when looking at the linearized dynamics around these semiclassical configurations.

\

Another point to note is the following. It is not obvious, at this stage, that the entire classical geometrodynamics has to be looked for in the mean field equations for the background GFT field selected. It is well possible that some (if not all) continuum geometric degrees of freedom are encoded in the GFT fluctuations around a given background field, e.g. one corresponding to a continuum but degenerate geometry, and that the latter has to be carefully chosen (or constrained by the GFT dynamics) so to admit a geometric re-writing and understanding {\it of the former}. 

Last, we note that our mean field equations gave also constraints on the relative value of the semi-classical parameter $t$, entering the definition of the coherent states chosen as mean field, and the GFT coupling constant $\lambda$. This coupling constant, although presently not well-understood, has been linked from various points of view to the cosmological constant \cite{iogft,laurentgft,iogft2,ACH}, and can be in general guessed to be related to the coupling constants of the emerging geometrical theory (be it General relativity of some modification of it). For example, in the case of 3d GFT, if this is something resembling 3d gravity, the obvious candidate would be the dimensionless quantity $G_{N} \Lambda$. While this is obviously only a speculation at present, we should also note that the same type of relation has been obtained in analog gravity models based on Bose condensates, where the semiclassicality (small depletion factor) was related to small values of $G_{N}\Lambda$ emerging in the Newtonian limit \cite{BECdyn}.  

Beyond the hydrodynamic (mean field) equations for the geometric order parameters, we have also extracted the effective dynamics for GFT perturbations around the chosen background, and pointed out some of their general properties, in particular the way they encode the microscopic dynamics and the differences between its spin foam expression and the original spin foam model. Finally we have discussed at some length the role of symmetries, in particular the recently identified GFT diffeomorphism symmetry, in the emergent effective dynamics, as compared to that in the fundamental GFT model.

\

Much more work remains to be done, both to develop and improve the results we have obtained, and to test if the path we are suggesting towards the solution of the problem of the continuum in quantum gravity is the correct one. We mention here a few lines of further research. 

The first focus should be the identification of the relevant macroscopic ground state for GFT hydrodynamics. Our choice is the most natural one from the point of view of LQG and spin foam models, as it corresponds to the semi-classical states used in both contexts for approximating macroscopic classical geometries from the pre-geometric data labeling kinematical LQG and SF boundary states. However, already at this level other choices are certainly possible, and the above states can be criticized on the grounds of being purely kinematical and possibly unstable (and  thus not truly semi-classical), when dynamics is taken into account. This seems to be confirmed by our analysis of their behaviour under the GFT dynamics. A different type of criticism comes from taking seriously the GFT framework as a second quantization of LQG (and thus of spin foam models) and the condensed matter analogy. If GFT are understood as a second quantized framework for spin network states or simplices \cite{iogft,laurentgft,iogft2,gftreview}, with LQG or simplicial wave functions in turn interpreted as many-particle states, then the semi-classical LQG states whose building block we used as background configuration for GFT dynamics in mean field expansion (and providing our geometric order parameters) are, in a sense, \lq\lq too semiclassical\rq\rq. They correspond, in fact, to quantum many-particle states such that each particle (spin network vertex or fundamental simplex) is individually semi-classical. Each vertex state is then given, as we have seen, by a product of wave functions associated to its links, and each of them is a semi-classical state for the corresponding quantum degrees of freedom. From the point of view of condensed matter, then, this means choosing as macroscopic configuration a state in which all the fundamental atoms are behaving semi-classically. In particular, this means that such states approximate very well a certain type of (kinematical) classical observables, but not those observable (\lq\lq extensive\rq\rq ones) that depend additively on the number of links and vertices of the spin network graph. Examples of such observables are most geometric observables like areas or volumes. It is well-possible, and even likely, that the origin of the correct classical dynamics of spacetime and the very emergence of continuum spacetime structures in Quantum Gravity is due to {\it quantum} properties of the underlying building blocks of quantum space, and thus captured by states for such building blocks that are {\it collectively} semi-classical, but individually highly quantum. Once more, an example of this type of behaviour is that of quantum liquids and of Bose condensates in particular. This example would suggest that the relevant vacuum state in the GFT context should be a second quantized coherent state, not a first quantized one (corresponding to LQG coherent states). This gives a further motivation for developing a second quantized picture and Fock structure for GFT states, which is the prerequisite for constructing such vacuum states.

A detailed analysis is needed for extracting the full geometric of the GFT-induced equations for the order parameters (classical phase space data) we have obtained, and of the effective spin foam dynamics for GFT perturbations. As for the first, a first step would be to try to separate the equations for the $SL(2,\mathbb{C})$ group elements into equations for their $\mathfrak{su}(2)$ and $\SU$ components, given that they have the classical interpretation of $B$ field and connection for the underlying BF theories. Assuming that our ansatz for the relevant vacuum state is the correct one, the final goal would be to establish a complete dictionary the GFT dynamics in mean field approximation, i.e. the GFT hydrodynamics, and the emergent classical theory (including GR). We are quite far from this goal yet, but an important asset is probably going to be the (non-commutative) metric representation of GFT, introduced in \cite{ioaristide}, that brings the geometric content of GFT dynamics to the forefront, both at the level of the action and of the Feynman amplitudes. In particular, for the purpose of extracting the geometric content of the GFT mean field equations, it will be useful to develop a (non-commutative) metric representation of the LQG coherent states and of our chosen GFT vacuum. Work on this is in progress. Similarly, the elucidation of the geometric content of the effective dynamics of GFT perturbations will benefit from such reformulation. Prior to this, however, it is important to study in more depth the physical interpretation of the perturbations themselves, to clarify if they carry gravitational/geometric degrees of freedom or if they should rather be interpreted as emergent matter \cite{emergentmatterGFT, eterawinston, ioeteraflorian}. Clearly, their interpretation depends strongly on the interpretation of the GFT vacuum chosen, which may induce, as in analog gravity condensed matter models \cite{analogreview, volovik, emergentmatterGFT}, a background geometry to  which the perturbations couple.
 
Finally, it is clear that the analysis presented in this paper, and the above-mentioned further steps, should be carried out in the physically relevant case of four spacetime dimension, and thus for 4d gravity GFT models. This on the one hand requires a better understanding of them than the one we have at present, and on the other hand it will contribute to such better understanding and development.

\

In any case, we believe that the first steps we have taken in this work indicate a path, from the microscopic picture of quantum space provided by group field theories (and other related approaches to Quantum Gravity) to the macroscopic description of the same consolidated in General Relativity, that is worth following further.

\acknowledgments 
We would like to thank B. Bahr and R. Pereira for useful discussions.
DO gratefully acknowledges financial support from the A. von Humboldt Stiftung through a Sofja Kovalevskaja Prize.
 
 \appendix

\section{$SL(2,\mathbb{C})$ matrix algebra} \label{appendixslc}
For convenience, we recall some basic properties of the $SL(2,\mathbb{C})$ matrices, and in particular the polar decomposition. Consider $M \in SL(2,\mathbb{C})$. There is a unique decomposition of it in terms of a $SU(2)$ matrix $U$ and a positive hermitian matrix with unit determinant $P$, $M=UP$. Consider the matrix:
\begin{equation}
{Q} = M^{\dagger}M
\end{equation}
It is a hermitian matrix, and hence it can be diagonalized via a unitary matrix $W$, such that:
\begin{equation}
Q = W D W^{-1},
\end{equation}
with $D$ a diagonal matrix. It is immediate to realize that this matrix is positive, \ie that the eigenvalues are (strictly) positive. This is due to the fact that the (diagonal) matrix elements of $Q$ can be related to norms of vectors once one recognizes that:
\begin{equation}
\langle x | Q |x \rangle = \langle x | M^{\dagger} M |x \rangle = || \,M |x\rangle \,||^{2}.
\end{equation}

Define the square root of $D$ to be the positive diagonal matrix $C$ such that $C^{2}=D$. 
With this matrix, define the square root $P$ of the matrix $Q$ to be the hermitian matrix
\begin{equation}
P = W C W^{-1}
\end{equation}
Given that $M$ is nonsingular, also $C$ must be nonsingular. Hence, it will admit an inverse, $C^{-1}$. Let us consider, then, the matrix:
\begin{equation}
U=M P^{-1} 
\end{equation}
Let us prove that it is unitary. Consider:
\begin{equation}
  UU^{\dagger} = M(P^{-1})^{2} M^{\dagger} = M (M^{\dagger}M)^{-1} M^{\dagger} = \mathbb{I} 
\end{equation}
This proves that $u \in U(2)$.
By keeping track of the signs of the determinants, one finds that $U \in SU(2)$.

Being a Hermitian matrix with unit determinant, belonging to $SL(2,\mathbb{C})$, the matrix $P$ admits the following representation:
\begin{equation}
P= \exp(e^{i}\sigma_{i}),
\end{equation}
with $\sigma_{i}$ the Pauli matrices\footnote{Notice that these are Hermitian matrices. Anti-Hermitian matrices are sometimes used in the LQG literature.} and $e^{i}$ real numbers (Einstein convention is assumed).

These matrices have a nice transformation property. Consider:
\begin{equation}
a = \exp(e^{i}\sigma_{i}), \qquad a' = w a w^{\dagger},
\end{equation}
with $w$ a $SU(2)$ matrix. Then:
\begin{equation}
a' = w \exp(e^{i}\sigma_{i}) w^{\dagger} = \exp(e^{i}w \sigma_{i}w^{\dagger}).
\end{equation}
However, since $w\sigma_{i}w^{\dagger}$ is a Hermitian, traceless matrix, it can be written as:
\begin{equation}
w\sigma_{i}w^{\dagger} = R_{i}^{j}(w) \sigma_{j},
\end{equation}
where $R^{i}_{j}(w)$ are just the rotation coefficients in the Lie algebra. Hence:
\begin{equation}
a' = \exp(e'^{i} \sigma_{i}), 
\qquad e'^{i} = R^{i}_{j}(w) e^{j}.
\end{equation}

\section{The action for 3D GFT} \label{appendixGFT}
It is worth to spend some words on a technical point that has been mentioned only briefly, \ie the construction of the interaction term needed for the construction of the GFT model for three dimensional theories.
We are going to discuss the issue step by step, in order to clarify some ambiguities that arise from a detailed analysis of the theory.

The starting point is an oriented tetrahedron. The four faces are oriented in such a way that their sides are ordered, respectively, $123$, $156$, $264$ and $453$ (see picture \ref{tetra}).
The desired GFT must be adapted to this combinatorial structure: the field arguments must be ordered in the appropriate way to respect the induced orientation on the sides of each face, and the gluings of the faces to form the tetrahedron itself. 

In building the model, we need a field $\phi_{123}$, which is not invariant under generic permutations of the arguments. Indeed, if this were the case, it would be impossible to encode any sort of orientation of the faces, and the perturbative expansion would include a larger class of simplicial complexes than the orientable ones. For the moment, we will assume no invariance at all under any form of permutation of indices. We will comment later on the possibility of having invariance under even permutations (\ie those respecting the orientation).

The kinetic term, providing the gluing of the faces of adjacent tetrahedra, in the perturbative expansion, must be such that triangles are glued, with normals pointing in opposite directions.
The natural candidate, then, is:
\begin{equation}
\phi_{123} \phi_{321},
\end{equation}
where the second field has arguments obtained via an odd permutation. Even here, we could consider seemingly equivalent alternatives:
\begin{equation*}
\phi_{123}\phi_{321};
\qquad \phi_{123} \phi_{213};
\qquad  \phi_{123} \phi_{132}.
\end{equation*}
It is worth mentioning what happens when varying the kinetic term, for each choice:
\begin{equation}
\frac{\delta }{\delta \phi_{abc}} \int (dg)^{3} \phi_{123} \phi_{321} = 2\phi_{cba} 
\end{equation}
\begin{equation}
\frac{\delta }{\delta \phi_{abc}} \int (dg)^{3} \phi_{123} \phi_{213} = 2\phi_{bac} 
\end{equation}
\begin{equation}
\frac{\delta }{\delta \phi_{abc}} \int (dg)^{3} \phi_{123} \phi_{132} = 2 \phi_{acb} 
\end{equation}
Therefore, the kinetic term alone will induce slightly different terms in the equation of motion, and hence a choice has to be done. We have chosen to use the paring $\phi_{123}\phi_{321}$.
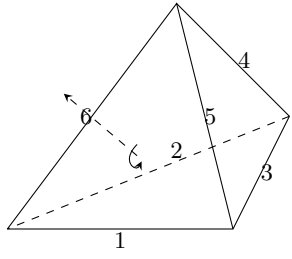
\begin{figure}
\begin{tikzpicture}[scale=1.5, >=stealth]
%\draw[step= 0.5, very thin] (0,0) grid (3,2);
\draw (0,0) -- (2,0) -- (2.5,1);
\draw[style=dashed] (2.5,1) -- (0,0);
\draw (0,0) -- (1.5,2) -- (2,0);
\draw (1.5,2) -- (2.5,1);
%\draw[style=dotted] (0,0) -- (1.7,1);
\draw[style=dashed,->](1.15,0.65) -- (0.5,1.2);
\draw[->](1.15,0.75) .. controls (1.,0.6)  and (1.14,0.5) .. (1.19,0.6);
\path (1,-0.1) node {$1$};
\path (2.3,0.5) node {$3$};
\path (1.5,0.7) node {$2$} ;
\path (1.8, 1) node {$5$};
\path (0.7, 1) node {$6$};
\path (2.1, 1.5) node {$4$};
\end{tikzpicture}
\caption{Oriented tetrahedron, with labeling of the edges.}
\label{tetra}
\end{figure}

The interaction term suffers from the same kind of ambiguities. Indeed, we can construct several inequivalent interaction terms, corresponding, naively, to the same combinatorial structure of the tetrahedron, but with rather different properties.
The first vertex we consider is:
\begin{equation}
\int \phi_{123} \phi_{156} \phi_{426} \phi_{453} (dg)^{6}.
\end{equation}
It is the one that we have considered in the paper. However, it is easy to see that we can generate a certain number of other operators by reshuffling the group elements in each field by means of an even permutation:
\begin{equation}
\int \phi_{123} \phi_{\pi_{1} \pi_{5} \pi_{6}}\phi_{\pi'_{4} \pi'_{2} \pi'_{6}}\phi_{\pi''_{4} \pi''_{5} \pi''_{3}} (dg)^{6}.
\end{equation}
To see that these vertices lead, in general, to different systems, we examine the first variation of this term.

\begin{equation}
\frac{\delta}{\delta{\phi_{abc}}}\int \phi_{123} \phi_{\pi_{1} \pi_{5} \pi_{6}}\phi_{\pi'_{4} \pi'_{2} \pi'_{6}}\phi_{\pi''_{4} \pi''_{5} \pi''_{3}} (dg)^{6}=
\end{equation}
\begin{eqnarray}
= \int \left\{ \delta^{a}_{1}\delta^{b}_{2}\delta^{c}_{3} \phi_{\pi_{1} \pi_{5} \pi_{6}}\phi_{\pi'_{4} \pi'_{2} \pi'_{6}}\phi_{\pi''_{4} \pi''_{5} \pi''_{3}} + \right. \\
\delta^{a}_{\pi_{1}}\delta^{b}_{\pi_{5}}\delta^{c}_{\pi_{6}} \phi_{1 23}\phi_{\pi'_{4} \pi'_{2} \pi'_{6}}\phi_{\pi''_{4} \pi''_{5} \pi''_{3}} + \\
\delta^{a}_{\pi'_{4}}\delta^{b}_{\pi'_{2}}\delta^{c}_{\pi'_{6}} \phi_{1 23}\phi_{\pi_{1} \pi_{5} \pi_{6}}\phi_{\pi''_{4} \pi''_{5} \pi''_{c}} + \\
\left.\delta^{a}_{\pi''_{4}}\delta^{b}_{\pi''_{5}}\delta^{c}_{\pi''_{3}} \phi_{1 23}\phi_{\pi'_{4} \pi'_{b} \pi'_{6}}\phi_{\pi_{1} \pi_{5} \pi_{6}} \right\} .
\end{eqnarray}

To uncover the structure of this term, we need to use the fact that the permutations form a group. We will need to use the inverse permutations, to be denoted by the letters $\omega,\omega',\omega''$:
\begin{equation}
\omega(\pi(a)) = a
\end{equation}

Therefore:
\begin{itemize}
\item The first contribution can be easily derived. It reads:
\begin{equation}
\phi_{\pi_{a} \pi_{5} \pi_{6}}\phi_{\pi'_{4} \pi'_{b} \pi'_{6}}\phi_{\pi''_{4} \pi''_{5} \pi''_{c}}
\end{equation}
\item $\delta^{a}_{\pi_{1}}\delta^{b}_{\pi_{5}}\delta^{c}_{\pi_{6}}$ implies:
\begin{equation}
g_{1} = g_{\omega(a)}; \qquad g_{5} = g_{\omega(b)}; \qquad g_{6} = g_{\omega(c)}, 
\end{equation}
so that the second contribution reads
\begin{equation}
 \phi_{\omega(a) 2 3 }\phi_{\pi'_{4} \pi'_{2} \pi'(\omega(c))}\phi_{\pi''_{4} \pi''(\omega(b)) \pi''_{3}} 
\end{equation}
\end{itemize}
Comparing this two terms, it is immediate to realize that they are distinct, in general: $g_{a},g_{b},g_{c}$ can appear in different positions among the arguments of the three fields. To see this explicitly, consider the case in which $\pi', \pi ''$ are just the identity. We would get the two terms:
\begin{equation}
\phi_{\pi_{a} \pi_{5} \pi_{6}}\phi_{4 b 6}\phi_{45c}, \qquad \phi_{\omega(a) 2 3 }\phi_{4 2 \omega(c)}\phi_{4 \omega(b)3}.
\end{equation}
For this to be the same term, we would need that also $\pi$ is the identity. This is the case for the term we have used to analyze the heat kernels within GFT.

This brief analysis shows how there is an ambiguity in the construction of the interaction term. However, this ambiguity is not only formal, but results in different equations of motion, \ie in different physical content of the theory. Of course, this will influence the partition function obtained from different choices, and, most importantly, the correlation functions of the models, given that they will obey different Schwinger--Dyson equations.

It is worth mentioning, however, that if we find a particular solution to the EOM with a particular choice of orderings, and this solution is invariant under general permutation of the group variables, then it will be a solution for the EOM obtained from any choice of orderings in the construction of the kinetic term and the interaction term.

\section{Useful facts about $\SU$ representations} \label{appendixreps}
It is convenient here to reproduce some basic facts about representations of $SU(2)$. For more details we will refer to \cite{vilenkin}.
The representation matrices that we need, $t^{l}_{mn}$, are labeled by a non-negative half integer $l$, and the convention on the matrix indices is that they run from $-l$ to $+l$. In terms of $SL2c$ matrices they have the following expressions:
\begin{equation}
t^{l}_{mn} \left( \begin{array} {cc}
\alpha & \beta \\
\gamma & \delta
\end{array} \right) = \alpha^{-(m+n)} \beta^{l+n} \gamma^{l+m}\sum_{j=M}^{N} C(l,m,n,j) \left( \frac{\alpha \delta}{\gamma \beta} \right)^{j},
\end{equation}
where the complex entries of the matrices are constrained to obey $\alpha\delta-\beta\gamma=1$, from the definition of $SL2C$, the coefficient appearing in the sum is defined to be
\begin{equation}
C(l,m,n,j)= \frac{\sqrt{(l+m)!(l-m)!(l+n)!(l-n)!}}{j! (j-m-n)!(l+n-j)!(l+m-j)!},
\end{equation}
and, finally, the sum is over the integers between:
\begin{equation}
M = \max (0,m+n), \qquad N= \min(l+m,l+n).
\end{equation}
These matrices reduce to the matrix representations of all the unitary, irreducible representations of $SU(2)$ when we restrict the group elements to belong to this subgroup of $SL2C$. 

The definition allows us to give a simple derivation of a result that is often needed in the manipulations discussed in this paper. It is rather straightforward to derive the following results:
\begin{equation}
t^{l}_{mn} \left( \begin{array} {cc}
\alpha & -\beta \\
-\gamma & \delta
\end{array} \right) = (-1)^{l+m}(-1)^{l+n} t^{l}_{mn} \left( \begin{array} {cc}
\alpha & \beta \\
\gamma & \delta
\end{array} \right),
\end{equation}
and
\begin{equation}
t^{l}_{mn} \left( \begin{array} {cc}
\delta & \beta \\
\gamma & \alpha
\end{array} \right) = t^{l}_{-n -m} \left( \begin{array} {cc}
\alpha & \beta \\
\gamma & \delta
\end{array} \right)
\end{equation}
Combining these two results, we can prove that:
\begin{equation}
t^{l}_{mn} (g) = (-1)^{l+m}(-1)^{l+n} t^{l}_{-n -m}(g^{-1}),
\end{equation}
for any element $g \in SLC$.
This is the key result that we need to prove:
\begin{equation}
\int_{SU(2)} dg \, t^{l}_{mn}(g) t^{l'}_{m'n'}(g) = \lj{l}{m}{m'} \lj{l}{n}{n'} \frac{\delta^{l l'}}{d_{l}}, 
\end{equation}
where we have defined the following object
\begin{equation}
\lj{l}{m}{m'} = \delta_{m,-m'} (-1)^{l+m'}.
\end{equation}


\begin{thebibliography}{99}


\bibitem{libro} D.~Oriti, Ed.,
  {\it Approaches to quantum gravity: Toward a new understanding of space, time
  and matter}
Cambridge, UK: Cambridge Univ. Pr. (2009).

\bibitem{thomas}
T.~Thiemann, {\it Modern canonical quantum general relativity}
Cambridge University Press (2007).

\bibitem{carlo} C.~Rovelli,
  {\it Quantum Gravity},  Cambridge, UK: Univ. Pr. (2004).

\bibitem{RC} H. Hamber, [arXiv:0704.2895 [hep-th]].

\bibitem{DT} R. Loll, Living Rev. Rel. 1, 13 (1998), [arXiv: gr-qc/9805049]; J. Ambjorn, J. Jurkiewicz, R. Loll, Phys.\ Rev.\ D \textbf{72}, 064014, (2005), [arXiv: hep-th/0505154].

\bibitem{review} D. Oriti,  Rept. Prog. Phys. \textbf{64}, 1489, (2001), [arXiv: gr-qc/0106091].

\bibitem{alex} A. Perez, Class. Quant. Grav. \textbf{20}, R43, (2003), [arXiv: gr-qc/0301113].

\bibitem{ST} J. Polchinski, {\it String Theory}, Cambridge University Press (1998); M. B. Green, J. H. Schwarz, E. Witten, {\it Superstring Theory}, Cambridge University Press (1988).

\bibitem{CS} J. Henson, in \cite{libro}, pp. 393-413, [arXiv: gr-qc/0601121].

\bibitem{topos} C. J. Isham, J. Butterfield, Found. Phys. 30, 1707-1735 (2000), [gr-qc/9910005].

\bibitem{NCG} G. Landi, [arXiv hep-th/9701078]; A. Connes, {\it Non-commutative Geometry}, Academic Press (1994); S. Majid, {\it Foundations of Quantum Group Theory}, Cambridge University Press (1995); S. Majid, J. Math. Phys. 41, 3892 (2000), [arXiv: hep-th/0006167].

\bibitem{iogft} D.~Oriti, in \cite{libro} pp. 310-331, [arXiv: gr-qc/0607032].

\bibitem{iogft2} D.~Oriti, in {\it Quantum Gravity}, B. Fauser, J. Tolksdorf and E. Zeidler, eds., Birkhaeuser, Basel, (2007), [arXiv: gr-qc/0512103].

\bibitem{laurentgft} L.~Freidel,
  Int.\ J.\ Theor.\ Phys.\  {\bf 44}, 1769 (2005)
  [arXiv: hep-th/0505016].

\bibitem{gftreview} D.~Oriti, in {\it Foundations of space and time\rq\rq} , G. Ellis, J. Murugan Eds, Cambridge, UK : Univ. Pr (2011)  

\bibitem{mm} F.~David,  Nucl. Phys. B257, \textbf{45} (1985); P.~Ginsparg, [arXiv: hep-th/9112013]; P.~Di Francesco, P.~H.~Ginsparg and J.~Zinn-Justin,
  Phys.\ Rept.\  {\bf 254}, 1 (1995)
  [arXiv: hep-th/9306153].

\bibitem{ioaristide} A. Baratin, D. Oriti, Phys.\ Rev.\ Lett.\  {\bf 105} (2010) 221302
  [arXiv:1002.4723 [hep-th]].

\bibitem{carlomike} M.~Reisenberger and C.~Rovelli,
  ``Spin foams as Feynman diagrams,''
  [arXiv: gr-qc/0002083].
  
\bibitem{coherentLQG}
  T.~Thiemann,
  Class.\ Quant.\ Grav.\  {\bf 18}, 2025 (2001)
  [arXiv:hep-th/0005233];
  T.~Thiemann and O.~Winkler,
  Class.\ Quant.\ Grav.\  {\bf 18}, 2561 (2001)
  [arXiv:hep-th/0005237];
  T.~Thiemann and O.~Winkler,
  Class.\ Quant.\ Grav.\  {\bf 18}, 4629 (2001)
  [arXiv:hep-th/0005234];
  T.~Thiemann and O.~Winkler,
  Class.\ Quant.\ Grav.\  {\bf 18}, 4997 (2001)
  [arXiv:hep-th/0005235];
 H.~Sahlmann, T.~Thiemann and O.~Winkler,
  Nucl.\ Phys.\  B {\bf 606} (2001) 401
  [arXiv:gr-qc/0102038];
  B.~Bahr and T.~Thiemann,
  Class.\ Quant.\ Grav.\  {\bf 26} (2009) 045011
  [arXiv:0709.4619 [gr-qc]];
   B.~Bahr and T.~Thiemann,
  Class.\ Quant.\ Grav.\  {\bf 26} (2009) 045012
  [arXiv:0709.4636 [gr-qc]].
 
\bibitem{fotini} T. Konopka, F. Markopoulou, S. Severini, Phys. Rev. D 77, 104029 (2008), [arXiv:0801.0861 [hep-th]].

\bibitem{fotini2} D. Kribs, F. Markopoulou, [gr-qc/0510052].

\bibitem{carlograviton} E.~Bianchi, L.~Modesto, C.~Rovelli and S.~Speziale,
  Class.\ Quant.\ Grav.\  {\bf 23}, 6989 (2006)
  [arXiv:gr-qc/0604044].
\bibitem{graviton2}
  C.~Rovelli,
  Phys.\ Rev.\ Lett.\  {\bf 97}, 151301 (2006)
  [arXiv:gr-qc/0508124].
  
\bibitem{SFcosmology} E. Bianchi, C. Rovelli, F. Vidotto, [arXiv:1003.3483 [gr-qc]].

\bibitem{gftfluid} 
  D.~Oriti,
  [arXiv:0710.3276 [gr-qc]].
\bibitem{hu}
  B.~L.~Hu,
  J.\ Phys.\ Conf.\ Ser.\  {\bf 174}, 012015 (2009)
  [arXiv:0903.0878 [gr-qc]]; B. L. Hu, Int. J. Theor. Phys. 44 (2005) 1785-1806,
[arXiv:gr-qc/0503067]; B. L. Hu,
[arXiv:gr-qc/9511076].  

\bibitem{analogreview}
  C.~Barcelo, S.~Liberati and M.~Visser,
  Living Rev.\ Rel.\  {\bf 8}, 12 (2005)
  [arXiv:gr-qc/0505065].

\bibitem{volovik}
  G.~E.~Volovik,
 {\it The Universe in a helium droplet,}
  Int.\ Ser.\ Monogr.\ Phys.\  {\bf 117}, 1 (2006); G. E. Volovik, rapporteur article for Proceedings of MG11, session `Analog Models of and for General
Relativity', arXiv:gr-qc/0612134; G. E. Volovik, JETP Lett. 82
(2005) 319-324; Pisma Zh.Eksp.Teor.Fiz. 82 (2005) 358-363,
[arXiv:gr-qc/0505104]; G. E. Volovik, Phys. Rept. 351 (2001)
195-348, [arXiv:gr-qc/0005091].


\bibitem{GRhydro} B.~L.~Hu,
[arXiv:gr-qc/9607070].

\bibitem{GRthermo}  T.~Padmanabhan,
  Rept.\ Prog.\ Phys.\  {\bf 73}, 046901 (2010)
  [arXiv:0911.5004 [gr-qc]].
  
  \bibitem{jacobson}
  T.~Jacobson,
  Phys.\ Rev.\ Lett.\  {\bf 75}, 1260 (1995)
  [arXiv:gr-qc/9504004].

  \bibitem{eling}
  C.~Eling, R.~Guedens and T.~Jacobson,
  Phys.\ Rev.\ Lett.\  {\bf 96}, 121301 (2006)
  [arXiv:gr-qc/0602001].

\bibitem{Eling2}
  C.~Eling,
  JHEP {\bf 0811}, 048 (2008)
  [arXiv:0806.3165 [hep-th]].

\bibitem{goffredo}
  G.~Chirco, C.~Eling and S.~Liberati,
  Phys.\ Rev.\  D {\bf 82}, 024010 (2010)
  [arXiv:1005.0475 [hep-th]].

 \bibitem{BECdyn}
  F.~Girelli, S.~Liberati and L.~Sindoni,
  Phys.\ Rev.\  D {\bf 78}, 084013 (2008)
  [arXiv:0807.4910 [gr-qc]];  S.~Liberati, F.~Girelli and L.~Sindoni,
 [arXiv:0909.3834 [gr-qc]].
 \bibitem{lor} L.~Sindoni, Phys.\ Rev.\ D {\bf 83} , 024022 (2011) [arXiv:1011.4411 [gr-qc]].

\bibitem{GFTrenorm} 
  L.~Freidel, R.~Gurau and D.~Oriti,
  Phys.\ Rev.\  D {\bf 80} (2009) 044007
  [arXiv:0905.3772 [hep-th]];  J. Magnen, K. Noui, V. Rivasseau and M. Smerlak, Class. Quant. Grav. \textbf{26}, 185012 (2009), [arXiv:0906.5477]. 

\bibitem{eterawinston}  W.~J.~Fairbairn and E.~R.~Livine,
  Class.\ Quant.\ Grav.\  {\bf 24}, 5277 (2007)
  [arXiv:gr-qc/0702125].
\bibitem{ioeteraflorian}F.~Girelli, E.~R.~Livine and D.~Oriti,
  Phys.\ Rev.\  D {\bf 81} (2010) 024015
  [arXiv:0903.3475 [gr-qc]].
  
\bibitem{emergentmatterGFT}
D. Oriti,
J.Phys.Conf.Ser.174:012047,2009, arXiv:0903.3970 [hep-th].

\bibitem{BEC} C.~J.~Pethick and H.~Smith,
{\em Bose-Einstein Condensation in Dilute Gases,}
Ed. Cambridge University Press, Cambridge, U.K. 2002.

\bibitem{castindum}
C.~W.~Gardiner, Phys.\ Rev.\ A {\bf 56} (1997), 1414-1423 [arXiv:quant-ph/9703005];
Y.~Castin and R. Dum, Phys.\ Rev. \ A {\bf 57} (1998), 3008-3021.

\bibitem{camporesi} R.~Camporesi,
  Phys.\ Rept.\  {\bf 196}, 1 (1990).

\bibitem{Hall} B. C. Hall, J. Mitchell, J. Math. Phys. 43, 1211-1236 (2002), Erratum-ibid.46, 059901 (2005),  [arXiv: quant-ph/0109086].

\bibitem{BF} D. Birmingham, M. Blau, M. Rakowski, G. Thompson, Phys. Rep. 209, 129 (1991).

\bibitem{Razvan} R. Gurau, [arXiv:0907.2582 [hep-th]].

\bibitem{Razvan2} % R. Gurau, [arXiv:0911.1945 [hep-th]]; R. Gurau, [arXiv:1006.0714 [hep-th]].
 R.~Gurau,
  Annales Henri Poincare {\bf 11} (2010) 565
  [arXiv:0911.1945 [hep-th]];
	R.~Gurau,
  Class.\ Quant.\ Grav.\  {\bf 27} (2010) 235023
  [arXiv:1006.0714 [hep-th]].

\bibitem{GFTdiffeo} A.~Baratin, F.~Girelli, D.~Oriti
  [arXiv:1101.0590 [hep-th]].

\bibitem{boulatov}
  D.~V.~Boulatov,
  Mod.\ Phys.\ Lett.\  A {\bf 7}, 1629 (1992)
  [arXiv:hep-th/9202074].

\bibitem{fluxes} A. Baratin, B. Dittrich, D. Oriti, J. Tambornino, [arXiv:1004.3450 [hep-th]].

 \bibitem{vincentcolored}J.~B.~Geloun, J.~Magnen and V.~Rivasseau,
  Eur.\ Phys.\ J.\  C {\bf 70} (2010) 1119
  [arXiv:0911.1719 [hep-th]].

\bibitem{vilenkin} N.~Ja.~Vilenkin and A.~U.~Klimyk, {\it Representation of Lie Groups and Special Functions,} Volume 1, Kluwer Academic Publishers, Dordrecht/Boston/London (1991).

\bibitem{simplicialDiffeos} B.Dittrich, [arXiv:0810.3594[gr-qc]].

\bibitem{PR3} L. Freidel, E. Livine, Class. Quant. Grav. 23, 2021(2006),  [arXiv: hep-th/0502106].

\bibitem{laurentmajid} L. Freidel, S. Majid, Class. Quant. Grav. \textbf{25}, 045006 (2008), [arXiv:hep-th/0601004].

\bibitem{karim} E. Joung, J. Mourad, K. Noui, J. Math. Phys. \textbf{50}, 052503 (2009), [arXiv:0806.4121 [hep-th]].

\bibitem{Tim} T. Koslowski, [arXiv:0709.3465 [gr-qc]].

\bibitem{Hanno} H. Sahlmann, Class. Quant. Grav. 27, 225007 (2010),  [arXiv:1006.0388 [gr-qc]].

\bibitem{EteraFloGFT} F.~Girelli and E.~R.~Livine,
	  Class.\ Quant.\ Grav.\  {\bf 27} (2010) 245018
  	[arXiv:1001.2919 [gr-qc]].

\bibitem{ACH} A. Ashtekar, M. Campiglia, A. Henderson,  Class. Quant. Grav. 27, 135020 (2010),  [arXiv:1001.5147 [gr-qc]].

\end{thebibliography}
\end{document}